\documentclass[useAMS,usenatbib]{tellus}

\usepackage{amsmath} 
\usepackage{amssymb} 
\usepackage{amsfonts}
\usepackage{graphicx,color}                 
\usepackage{subfigure}
\usepackage{lineno}
\usepackage{multirow}
\linenumbers

\begin{document}

\title[Ensemble Kalman filtering with residual nudging]
{Ensemble Kalman filtering with residual nudging}

\author[By X. Luo and I. Hoteit]{By Xiaodong Luo\thanks{Corresponding
author.\hfil\break e-mail: xiaodong.luo@iris.no}$^a$ and
Ibrahim Hoteit$^b$,} \affiliation{$^a$International Research Institute of Stavanger, 5008 Bergen, Norway \\
$^b$King Abdullah University of Science and Technology, Thuwal, 23955-6900, Saudi Arabia}


\maketitle
\begin{abstract}
Covariance inflation and localization are two important techniques that are used to improve the performance of the ensemble Kalman filter (EnKF) by (in effect) adjusting the sample covariances of the estimates in the state space. In this work an additional auxiliary technique, called residual nudging, is proposed to monitor and, if necessary, adjust the residual norms of state estimates in the observation space. In an EnKF with residual nudging, if the residual norm of an analysis is larger than a pre-specified value, then the analysis is replaced by a new one whose residual norm is no larger than a pre-specified value. Otherwise the analysis is considered as a reasonable estimate and no change is made. A rule for choosing the pre-specified value is suggested. Based on this rule, the corresponding new state estimates are explicitly derived in case of linear observations. Numerical experiments in the 40-dimensional Lorenz 96 model show that introducing residual nudging to an EnKF may improve its accuracy and/or enhance its stability against filter divergence, especially in the small ensemble scenario.
\end{abstract}

\section{Introduction}
The ensemble Kalman filter (EnKF) \citep{Anderson-ensemble,Bishop-adaptive,Burgers-analysis,Evensen-sequential,Hoteit2002,Houtekamer1998,Pham2001,Whitaker-ensemble} is an efficient algorithm for data assimilation in high dimensional systems. Because of its runtime efficiency and simplicity in implementation, it is receiving ever-increasing attentions from researchers in various fields. In many applications of the EnKF, due to limited computational resources, one is only able to run an EnKF with an ensemble size much smaller than the dimension of the state space. In such circumstances, problems often arise, noticeably on the quality of the sample covariances, including, for instance, rank-deficiency, underestimation of the covariance matrices \citep{Sacher2008-sampling,Whitaker-ensemble}, and spuriously large cross-variances between independent (or uncorrelated) state variables \citep{Hamill-distance}. To mitigate these problems, it is customary to introduce two auxiliary techniques, namely covariance inflation \citep{Anderson-Monte} and localization \citep{Hamill-distance}, to the EnKF. On the one hand, covariance inflation increases the estimated sample covariances in order to compensate for the effect of underestimation, which in fact increases the robustness of the EnKF in the sense of \citet{Luo2011_EnLHF}. On the other hand, covariance localization introduces a ``distance''-dependent tapering function to the elements of the sample covariances, and smooths out the spuriously large values in them. In addition, covariance localization also increases the ranks of the sample covariances \citep{Hamill2009}.

Both covariance inflation and localization are techniques that in effect adjust the sample covariances in the state space. Since data assimilation is a practice of estimation that incorporates information from both the state and observation spaces, it would be natural for one to make use of the information in the observation space to improve the performance of an EnKF.

In this study we propose such an observation-space based auxiliary technique, called residual nudging, for the EnKF. Here a ``residual'' is a vector in the observation space, and is defined as the projection of an analysis mean onto the observation space subtracted from the corresponding observation. In residual nudging our objective is to make the vector norm of the residual (``residual norm'' for short) no larger than a pre-specified value. This is motivated by the observation that, if the residual norm is too large, then the corresponding analysis mean is often a poor estimate. In such cases, it is better off to choose as the new estimate a state vector whose residual norm is smaller.

The method presented in this work is close to the idea of \citet{vanLeeuwen2010-nonlinear}, in which a nudging term is added to the particle filter so that the projections of the particles onto the observation space are drawn closer to the corresponding observation, and the particles themselves are associated with almost equal weights. By doing so, the modified particle filter can achieve remarkably good performance using only 20 particles in the chaotic 40-dimensional Lorenz-96 (L96) model \citep{Lorenz-optimal}, while traditional methods may need thousands of particles \citep{vanLeeuwen2010-nonlinear}. Other similar, residual-related, methods were also found in the literature, for examples, see \citet{Anderson2007,Anderson2009,Song2010-adaptive}. \citet{Anderson2007,Anderson2009} suggested adaptive covariance inflation schemes in the context of hierarchical ensemble filtering. There the inflation factor $\lambda$ is considered as a random variable (with a presumed initial prior distribution), and in effect adjusts the projection of the background (co)variances onto the observation space\footnote{In contrast, in residual nudging we are interested in adjusting the projection of the background mean. Comparison and/or combination of these two strategies will be deferred to future investigations.}. With an incoming observation, the prior distribution is updated to the posterior one based on Bayes' rule, while the residual affects the shape of the posterior distribution of $\lambda$. On the other hand, \citet{Song2010-adaptive} considered the idea of replacing an existing analysis ensemble member by a new one, in which the residual plays a role in generating the new ensemble member.

Our main purpose here is to use residual nudging as a safeguard strategy, with which the projections of state estimates onto the observation space, under suitable conditions, are guaranteed to be within a pre-specified distance to the corresponding observations. We will discuss how to choose the pre-specified distance, and construct the (possibly) new state estimates accordingly in case of linear observations. In this work, the ensemble adjustment Kalman filter (EAKF) \citep{Anderson-ensemble} is adopted for the purpose of demonstration, while the extension to other filters can be done in a similar way. Through numerical experiments in the L96 model, we show that, the EAKF equipped with residual nudging (EAKF-RN) is more robust than the normal EAKF. In addition, the accuracy of the EAKF-RN is comparable to, and sometimes (much) better than, that of the normal EAKF.

This work is organized as follows. Section \ref{sec:EAKF-RN} reviews the filtering step of the EAKF, introduces the concept of residual nudging, and discusses how it can be implemented in the EAKF. Section \ref{sec:numerical_example_KF} investigates the effect of residual nudging on the performance of the Kalman filter (KF) in a linear/Gaussian system, which aims to provide some insights of how residual nudging may affect the behaviour of an already optimal filter. Section \ref{sec:numerical_example} extends the investigation to the Lorenz 96 model, in which we examine the performance of the EAKF-RN in various scenarios, and compare it with the normal EAKF. Section \ref{sec:conclusion} discusses possible extensions of the current study and concludes the work.

\section{Ensemble Kalman filtering with residual nudging} \label{sec:EAKF-RN}

Suppose that at the $k$th assimilation cycle, one has a background ensemble $\mathbf{X}^b_k = \{ \mathbf{x}_{k,i}^b \}_{i=1}^n$ with $n$ members. The incoming observation $\mathbf{y}_k^o$ is obtained from the following observation system
\begin{linenomath*}
\begin{equation} \label{eq:observation_process}
\mathbf{y}_k = \mathbf{H}_k \mathbf{x}_k + \mathbf{v}_k \, ,
\end{equation}
\end{linenomath*}
where $\mathbf{H}_k$ is a matrix, and $\mathbf{v}_k$ is the observation noise, with zero mean and covariance $\mathbf{R}_k$. For convenience of discussion, we assume that the dimensions of $\mathbf{x}_k$ and $\mathbf{y}_k$ are $m_x$ and $m_y$, respectively, $m_y \leq m_x$, and $\mathbf{H}_k$ has full row rank.


\subsection{The filtering step of the ensemble adjustment Kalman filter with covariance inflation and localization} \label{subsec:filtering_EAKF}

We first summarize the filtering step of the EAKF with both covariance inflation and localization. For simplicity, here we only consider the scenario with constant covariance inflation and localization, and refer readers to, for example, \citet{Anderson2007,Anderson2009}, for the details of adaptive configuration of the EAKF. In the context of EAKF, it is assumed that the covariance $\mathbf{R}_k$ of the observation noise is a diagonal matrix, such that one can assimilate the incoming observation in a serial way. Following \citet{Anderson2007,Anderson2009}, we use a single scalar observation to demonstrate the assimilation algorithm in the EAKF. To this end, in this sub-section (only) we temporarily assume that the observation vector $\mathbf{y}_k \equiv y_k$ is a scalar random variable, with zero mean and variance $R_k$. The notation of the incoming observation thus becomes $y^o_k$, with the dimension $m_y = 1$. The algorithm description below mainly follows \citet{Anderson2007}.

Suppose that the $i$-th ensemble member $\mathbf{x}_{k,i}^b$ of $\mathbf{X}^b_k$ consists of $m_x$ elements $(\mathbf{x}_{k,i}^b)_j$ ($j = 1, \dotsb, m_x$) such that $\mathbf{x}_{k,i}^b = [(\mathbf{x}_{k,i}^b)_1,\dotsb,(\mathbf{x}_{k,i}^b)_{m_x}]^T$. Then the sample mean $\hat{\mathbf{x}}_k^b$ of $\mathbf{X}^b_k$ is $\hat{\mathbf{x}}_k^b = \sum\limits_{i=1}^n \mathbf{x}_{k,i}^b / n $. To introduce covariance inflation to the filter, suppose that $\Delta \mathbf{X}^b_k \equiv \{ \Delta \mathbf{x}_{k,i}^b: \Delta \mathbf{x}_{k,i}^b = \mathbf{x}_{k,i}^b - \hat{\mathbf{x}}_k^b \}_{i=1}^n$ is the ensemble of deviations with respect to $\mathbf{X}^b_k$, and $\lambda \geq 1$ the inflation factor, then the inflated background ensemble is $\mathbf{X}^{inf}_k \equiv \{ \mathbf{x}_{k,i}^{inf}: \mathbf{x}_{k,i}^{inf} = \hat{\mathbf{x}}_k^b + \sqrt{\lambda} \: \Delta \mathbf{x}_{k,i}^b \}_{i=1}^n$ \citep{Anderson2007,Anderson2009}. With covariance inflation, $\mathbf{X}^{inf}_k$ and $\mathbf{X}^{b}_k$ have the same mean, but the sample covariance of $\mathbf{X}^{inf}_k$ is $\lambda$ times that of $\mathbf{X}^{b}_k$. In what follows, we do not particularly distinguish background ensembles with and without covariance inflation through different notations. Instead, we always denote the background ensemble by $\mathbf{X}^{b}_k$, no matter whether it is inflated or not. One can tell whether a background ensemble is inflated by checking the value of $\lambda$, e.g., $\lambda = 1$ means no inflation, and $\lambda > 1$ with covariance inflation.

On the other hand, suppose that the projection of $\mathbf{X}^b_k$ onto the observation space is $\mathbf{Y}^b_k = \{ y_{k,i}^b: y_{k,i}^b = \mathbf{H}_k \mathbf{x}_{k,i}^b \}_{i=1}^n$, then one can compute the sample mean $\hat{y}^b_k$ and sample variance $\hat{p}^b_{yy,k}$ as
\begin{linenomath*}
\begin{equation} \label{eq:bg_prj_mean_cov}
\begin{split}
& \hat{y}^b_k = \dfrac{1}{n} \sum\limits_{i=1}^n y_{k,i}^b \, ,  \\
& \hat{p}^b_{yy,k} = \dfrac{1}{n-1} \sum\limits_{i=1}^n (y_{k,i}^b - \hat{y}^b_k)^2 \, .
\end{split}
\end{equation}
\end{linenomath*}
With the incoming observation $y^o_k$, one updates $\hat{y}^b_k$ and $\hat{p}^b_{yy,k}$ to their analysis counterparts, $\hat{y}^a_k$ and $\hat{p}^a_{yy,k}$, respectively, through the following formulae \citep[Eq. (3.2 - 3.3)]{Anderson2007}.
\begin{linenomath*}
\begin{equation} \label{eq:ana_prj_mean_cov}
\begin{split}
& \hat{p}^a_{yy,k} = [(\hat{p}^b_{yy,k})^{-1} + R_k^{-1}]^{-1} \, , \\
& \hat{y}^a_k =\hat{p}^a_{yy,k} [(\hat{p}^b_{yy,k})^{-1} \hat{y}^b_k + R_k^{-1} y^o_k] \, .
\end{split}
\end{equation}
\end{linenomath*}
Accordingly, one can update the projection $\mathbf{Y}^b_k$ to its analysis counterpart $\mathbf{Y}^a_k \equiv \{ y_{k,i}^a: y_{k,i}^a = y_{k,i}^b + \delta y_{k,i} \}_{i=1}^n$, where the increments $\delta y_{k,i}$ with respect to $y_{k,i}^b$ are given by
\begin{linenomath*}
\begin{equation} \label{eq:obv_increments}
\delta y_{k,i}  = \sqrt{\dfrac{\hat{p}^a_{yy,k}}{\hat{p}^b_{yy,k}}} \; (y_{k,i}^b - \hat{y}^b_k) + \hat{y}^a_k -  y_{k,i}^b \, .
\end{equation}
\end{linenomath*}
One can verify that the sample mean and covariance of $\mathbf{Y}^a_k$ are $\hat{y}^a_k$ and $\hat{p}^a_{yy,k}$, respectively. Also note the difference between the concepts of deviations and increments. For distinction we have used $\Delta$ to denote deviations, and $\delta$ increments.

After the above quantities are calculated, one proceeds to update the background ensemble $\mathbf{X}^b_k$ to the analysis one $\mathbf{X}^a_k \equiv \{ \mathbf{x}_{k,i}^a: \mathbf{x}_{k,i}^a = \mathbf{x}_{k,i}^b + \delta \mathbf{x}_{k,i}  \}_{i=1}^n$, where the increment $\delta \mathbf{x}_{k,i}$ with respect to the $i$-th background ensemble member $\mathbf{x}_{k,i}^b$ is an $m_x$ dimensional vector, i.e., $\delta \mathbf{x}_{k,i} = [(\delta \mathbf{x}_{k,i})_1,\dotsb,(\delta \mathbf{x}_{k,i})_{m_x}]^T$, where the $j$-th element $(\delta \mathbf{x}_{k,i})_j$ of $\delta \mathbf{x}_{k,i}$ is given by
\begin{linenomath*}
\begin{equation} \label{eq:state_increments}
(\delta \mathbf{x}_{k,i})_j  = (\hat{p}^j_{xy,k}/\hat{p}^b_{yy,k}) \delta y_{k,i} \, , j = 1, \dotsb, m_x \, ,
\end{equation}
\end{linenomath*}
with $\hat{p}^j_{xy,k}$ being the sample cross-variance between all the $j$-th elements of the ensemble members of $\mathbf{X}^b_k$, and the projection ensemble $\mathbf{Y}^b_k = \{ y_{k,i}^b \}_{i=1}^n$, i.e.,
\begin{linenomath*}
\begin{equation} \label{eq:eakf_cross_variance}
\hat{p}^j_{xy,k} = \dfrac{1}{n-1} \sum\limits_{i=1}^n [(\mathbf{x}_{k,i}^b)_j - (\hat{\mathbf{x}}_{k}^b)_j ] [y_{k,i}^b - \hat{y}^b_k] \, .
\end{equation}
\end{linenomath*}
With relatively small ensemble sizes, Eq.~(\ref{eq:eakf_cross_variance}) often results in spuriously large sample cross-variances \citep{Hamill-distance}. To tackle this problem, one may introduce covariance localization \citep{Hamill-distance} to the EAKF, in which the main idea is to multiply $\hat{p}^j_{xy,k}$ in Eq.~(\ref{eq:state_increments}) by a ``distance''-dependent tapering coefficient $\eta_{ij} \leq 1$ \citep{Anderson2007,Anderson2009}. We will discuss how to compute $\eta_{ij}$ in the experiments with respect to the L96 model.

After obtaining the analysis ensemble $\mathbf{X}^a_k$, one computes the analysis mean  $\hat{\mathbf{x}}_k^a = \sum\limits_{i=1}^n \mathbf{x}_{k,i}^a / n $ (analysis for short), and uses it as the posterior estimate of the system state. Propagating $\mathbf{X}^a_k$ forward through the dynamical model, a background ensemble at the next assimilation time is obtained, and a new assimilation cycle starts, and so on.

\subsection{Residual nudging} \label{sec:RN}

As will be shown later, the EAKF may suffer from filter divergence in certain circumstances, even when it is equipped with both covariance inflation and localization. To mitigate filter divergence, intuitively one may choose to adjust the estimate $\hat{\mathbf{x}}^a_k$ and move it closer toward the truth $\mathbf{x}_k^{tr}$. In practice, though, $\mathbf{x}_k^{tr}$ is normally unknown, thus it is infeasible to apply this state-space based strategy. In what follows, we introduce a similar, but observation-space based strategy, in which the main idea is to monitor, and, if necessary, adjust the residual norm of the estimate. For this reason we refer to this strategy as residual nudging.

By definition, the residual with respect to the analysis mean $\hat{\mathbf{x}}^a_k$ is $\hat{\mathbf{r}}_k^a \equiv \mathbf{H}_k \hat{\mathbf{x}}^a_k - \mathbf{y}_k^o$. We also define the 2-norm of a vector $\mathbf{z}$ as
\begin{linenomath*}
\begin{equation} \label{eq:2_norm}
\Vert \mathbf{z} \Vert_2 \equiv \sqrt{\mathbf{z}^T \mathbf{z}} \, .
\end{equation}
\end{linenomath*}
The objective in residual nudging is the following. We accept $\hat{\mathbf{x}}^a_k$ as a reasonable estimate if its residual norm $\Vert \hat{\mathbf{r}}_k^a \Vert_2$ is no larger than a pre-specified value, say, $\beta \sqrt{\text{trace}(\mathbf{R}_k)}$, with $\beta >0$ being called the noise level coefficient hereafter (the reason in choosing this pre-specified value will be explained soon). Otherwise, we consider $\hat{\mathbf{x}}^a_k$ a poor estimate, and thus find for it a replacement, say, $\tilde{\mathbf{x}}^a_k$, based on the estimate $\hat{\mathbf{x}}^a_k$ and the observation $\mathbf{y}_k^o$, so that the residual norm of $\tilde{\mathbf{x}}^a_k$ is no larger than $\beta \sqrt{\text{trace}(\mathbf{R}_k)}$. To this end, we stress that the assumption $m_y \leq m_x$ may be necessary in certain cases (see the discussion later). In this work we focus on the cases with $m_y \leq m_x$, which is true for many geophysical problems.

The objective of residual nudging can be achieved as follows. First of all, we compute a scalar $c_k \in [0,1]$, called the fraction coefficient hereafter (cf. Eq.~(\ref{eq:replacement_mixing}) later for the reason), according to the formula
\begin{linenomath*}
\begin{equation} \label{eq:fraction_coefficient}
c_k = \min(1, \beta \sqrt{\text{trace}(\mathbf{R}_k)}/\Vert \hat{\mathbf{r}}_k^a \Vert_2) \, ,
\end{equation}
\end{linenomath*}
where the function $\min(a,b)$ finds the minimum between the scalars $a$ and $b$. The rationale behind Eq.~(\ref{eq:fraction_coefficient}) is this: if $\Vert \hat{\mathbf{r}}_k^a \Vert_2 > \beta \sqrt{\text{trace}(\mathbf{R}_k)}$, then we need to multiply $\Vert \hat{\mathbf{r}}_k^a \Vert_2$ by a coefficient $c_k<1$ to reduce $\Vert \hat{\mathbf{r}}_k^a \Vert_2$ to the pre-specified value. Otherwise, we do nothing and keep $\Vert \hat{\mathbf{r}}_k^a \Vert_2$ as it is, which is equivalent to multiplying $\Vert \hat{\mathbf{r}}_k^a \Vert_2$ by $c_k = 1$.

Next, we construct a new estimate $\tilde{\mathbf{x}}^a_k$ by letting
\begin{linenomath*}
\begin{subequations} \label{eq:replacement}
\begin{align}
\label{eq:replacement_mixing} & \tilde{\mathbf{x}}^a_k = c_k  \, \hat{\mathbf{x}}^a_k + (1-c_k) \, \mathbf{x}^{o}_k \, , \\
\label{eq:pseudo_inversion} & \mathbf{x}^{o}_k = \mathbf{H}_k^T ( \mathbf{H}_k \mathbf{H}_k^T)^{-1} \mathbf{y}^{o}_k \ .
\end{align}
\end{subequations}
\end{linenomath*}
The term $\mathbf{H}_k^T ( \mathbf{H}_k \mathbf{H}_k^T)^{-1}$ in Eq.~(\ref{eq:pseudo_inversion}) is the Moore-Penrose generalized inverse of $\mathbf{H}_k$, such that $\mathbf{x}^{o}_k$ in Eq.~(\ref{eq:pseudo_inversion}) provides a least-square solution for the equation $\mathbf{H}_k \mathbf{x} = \mathbf{y}^{o}_k$ \citep[ch. 2]{Engl2000-regularization}. We refer to $\mathbf{x}^{o}_k$ as the observation inversion hereafter. With Eq.~(\ref{eq:replacement}), the new residual $\tilde{\mathbf{r}}^a_k = \mathbf{H}_k \tilde{\mathbf{x}}^a_k - \mathbf{y}^{o}_k = c_k \, \hat{\mathbf{r}}^a_k$, so that $ \Vert \tilde{\mathbf{r}}^a_k \Vert_2 = c_k \, \Vert \hat{\mathbf{r}}^a_k \Vert_2 \leq \beta \sqrt{\text{trace}(\mathbf{R}_k)}$ according to Eq.~(\ref{eq:fraction_coefficient}).

In residual nudging we only attempt to adjust the analysis mean $\hat{\mathbf{x}}^a_k$ of the EAKF, but not its covariance. To this end, let the analysis ensemble be $\mathbf{X}^a_k = \{ \mathbf{x}_{k,i}^a: \mathbf{x}_{k,i}^a = \hat{\mathbf{x}}_{k}^a + \Delta \mathbf{x}_{k,i}^a  \}_{i=1}^n$, where the deviations $\Delta \mathbf{x}_{k,i}^a = \mathbf{x}_{k,i}^a - \hat{\mathbf{x}}_{k}^a$. We then replace the original analysis mean $\hat{\mathbf{x}}^a_k$ by $\tilde{\mathbf{x}}^a_k$, and change the analysis ensemble to $\tilde{\mathbf{X}}^a_k = \{ \tilde{\mathbf{x}}_{k,i}^a: \tilde{\mathbf{x}}_{k,i}^a = \tilde{\mathbf{x}}^a_k + \Delta \mathbf{x}_{k,i}^a  \}_{i=1}^n$. Therefore, in comparison with the normal EAKF, the EAKF with residual nudging (EAKF-RN for short) just has additional steps in Eqs.~(\ref{eq:fraction_coefficient}) and(\ref{eq:replacement}), while all the other procedures remain the same. In doing so, residual nudging is compatible with both covariance inflation and localization.

\subsection{Discussion} \label{sec:RN discussion}
Choosing the pre-specified value in the form of $\beta \sqrt{\text{trace}(\mathbf{R}_k)}$ is motivated by the following consideration. Let $\mathbf{x}_k^{tr}$ be the truth such that $\mathbf{y}_k^o = \mathbf{H}_k  \mathbf{x}_k^{tr} +  \mathbf{v}_k$. Then $\tilde{\mathbf{r}}_k^a = \mathbf{H}_k \tilde{\mathbf{x}}^a_k - \mathbf{y}_k^o =\mathbf{H}_k (\tilde{\mathbf{x}}^a_k - \mathbf{x}_k^{tr}) - \mathbf{v}_k$, and by the triangle inequality,
\begin{linenomath*}
\begin{equation} \label{eq:residual_norm}
\Vert \tilde{\mathbf{r}}_k^a \Vert_2 \leq \Vert \mathbf{H}_k (\tilde{\mathbf{x}}^a_k - \mathbf{x}_k^{tr}) \Vert_2 + \Vert \mathbf{v}_k \Vert_2  \, .
\end{equation}
\end{linenomath*}
For a reasonably good estimate $\tilde{\mathbf{x}}^a_k$, we expect that the magnitude of $\mathbf{H}_k \tilde{\mathbf{x}}^a_k - \mathbf{H}_k \mathbf{x}_k^{tr}$ should not substantially exceed the observation noise level. On the other hand, we have $ (\mathbb{E} \Vert \mathbf{v}_k \Vert_2)^2 \leq \mathbb{E} \Vert \mathbf{v}_k \Vert_2^2 = \text{trace} (\mathbb{E} (\mathbf{v}_k \mathbf{v}_k^T) ) = \text{trace} (\mathbf{R}_k)$, thus the expectation $\mathbb{E} \Vert \mathbf{v}_k \Vert_2$ of the norm of the observation noise is (at most) in the order of $\sqrt{\text{trace} (\mathbf{R}_k)}$. One may thus use $\sqrt{\text{trace} (\mathbf{R}_k)}$ to characterize the noise level. By requiring that a reasonably good estimate have $\Vert \mathbf{H}_k (\tilde{\mathbf{x}}^a_k - \mathbf{x}_k^{tr}) \Vert_2$ in the order of $\sqrt{\text{trace} (\mathbf{R}_k)}$ (or less), one comes to the choice in the form of $\beta \sqrt{\text{trace}(\mathbf{R}_k)}$. The criterion in choosing the above threshold is very similar to that in certain quality control algorithms (called check of plausibility, see, for example \citealp{gandin1988complex}, for a survey), in which one is assumed to have prior knowledge about, say, the mean $\bar{y}_s$ and variance $\sigma_s$ of a scalar observation $y_{s}$. In quality control, $y_s$ is often assumed to be a Gaussian random variable, so that for a measured observation $y^o_s$, if the ratio $|y^o_s - \bar{y}_s|/\sigma_s$ is too large, then $y^o_s$ is discarded, or at least suspected \citep{gandin1988complex}. The main differences between residual nudging and quality control are the following. While quality control checks the plausibility of an incoming observation, residual nudging checks the plausibility of a state estimate, and suggests a replacement if the original state estimate does not pass the test. Moreover, as long as the $2$-norm is used, the expectation $\mathbb{E}\Vert \mathbf{v}_k \Vert_2^2$ is always $\text{trace}(\mathbf{R}_k)$, independent of the distribution of $\mathbf{v}_k$. This independence, on the one hand, implies that the inequality in (\ref{eq:residual_norm}), hence the threshold $\beta \sqrt{\text{trace}(\mathbf{R}_k)}$, holds without requiring the knowledge of the distribution of $\mathbf{H}_k (\tilde{\mathbf{x}}^a_k - \mathbf{x}_k^{tr})$. On the other hand, the absence of the knowledge of the distribution means that less statistical information is gained in choosing the threshold $\beta \sqrt{\text{trace}(\mathbf{R}_k)}$. For instance, one may not be able to assign a statistical meaning to $\beta \sqrt{\text{trace}(\mathbf{R}_k)}$, nor obtain a confidence (or significance) level in accepting (or rejecting) a state estimate. Finally, it is also possible for one to adopt another distance metric, e.g., the $1$- or $\infty$-norm, for which the inequality in (\ref{eq:residual_norm}) still holds. In such circumstances, the expectation, $\mathbb{E}\Vert \mathbf{v} \Vert_1^2$ or $\mathbb{E}\Vert \mathbf{v}_k \Vert_{\infty}^2$, may not be equal to $\text{trace}(\mathbf{R}_k)$ any more, so that one may need to choose a threshold different from $\beta \sqrt{\text{trace}(\mathbf{R}_k)}$. Despite the stated differences, we expect that residual nudging can be used in conjunction with observation quality control, although this is not pursed in the current study.

Even though the noise level coefficient $\beta$ in residual nudging is chosen to be time-invariant, the resulting fraction coefficient $c_k$ in general changes with time according to Eq. (\ref{eq:fraction_coefficient}). The coefficient $\beta$ affects how the new analysis $\tilde{\mathbf{x}}^a_k$ combines the original one $\hat{\mathbf{x}}^a_k$ and the observation inversion $\mathbf{x}_k^o$. This can be seen from Eqs.~(\ref{eq:fraction_coefficient}) and (\ref{eq:replacement_mixing}). Because $c_k \in [0,1]$,  the new analysis $\tilde{\mathbf{x}}^a_k$ in Eq.~(\ref{eq:replacement_mixing}) is a convex combination of  $\hat{\mathbf{x}}^a_k$ and $\mathbf{x}^{o}_k$, i.e., an estimate somewhere in-between the original estimate $\hat{\mathbf{x}}^a_k$ and the observation inversion $\mathbf{x}^{o}_k$, depending on the value of $c_k$. If one chooses a large value for $\beta$, or, if for a fixed $\beta$ the original residual norm $\hat{\mathbf{r}}_k^a$ is sufficiently small, then the fraction coefficient $c_k \rightarrow 1$ according to Eq.~(\ref{eq:fraction_coefficient}), thus $\tilde{\mathbf{x}}^a_k \rightarrow \hat{\mathbf{x}}^a_k$ according to Eq.~(\ref{eq:replacement_mixing}). Therefore $\tilde{\mathbf{x}}^a_k$ will be a good estimate if $\hat{\mathbf{x}}^a_k$ is so, but may not be able to achieve a good estimation accuracy when $\hat{\mathbf{x}}^a_k$ itself is poor. On the other hand, if one chooses a very small value for $\beta$, or, if for a fixed $\beta$ the original residual norm $\hat{\mathbf{r}}_k^a \rightarrow +\infty$ (e.g., with filter divergence), then $c_k \rightarrow 0$ and $\tilde{\mathbf{x}}^a_k \rightarrow \mathbf{x}^{o}_k$. In this case, the estimate $\tilde{\mathbf{x}}^a_k$ is calculated mainly based on the information content of the observation $\mathbf{y}^{o}_k$, and may result in a relatively poor accuracy. This is largely because of (1) the presence of the observation noise $\mathbf{v}_k$ in Eq.~(\ref{eq:observation_process}), and (2) the ignorance of the prior knowledge of the model dynamics. As a result, pushing the projection of state estimates very close to noisy observations may have some negative consequences. For instance, in geophysical applications, dynamical balances of the numerical models may not be honored so that the estimation errors may be relatively large. However, using $\mathbf{x}^{o}_k$ as the estimate may be a relatively safe (although conservative) strategy against filter divergence. In the sense of the above discussion, the choice of $\beta$ reflects the extent to which one wants to achieve the trade-off between a filter's potential accuracy and stability against divergence. This point is further demonstrated through some experiments later.

Some numerical issues related to the computation of the observation inversion $\mathbf{x}^{o}_k$ are discussed in order. One is the existence and uniqueness of the observation inversion. Under the assumptions that $m_y \leq m_x$ and that $\mathbf{H}_k$ is of full row rank, the observation inversion, as a solution of the equation $\mathbf{H}_k \mathbf{x} = \mathbf{y}^{o}_k$, does exist \citep[ch. 4]{Meyer2001}. Finding a concrete solution, however, is in general an under-determined problem, hence the solution is not unique unless $m_y = m_x$. This point can be seen as follows. When $m_y < m_x$, the null space $\mathbb{S}^N$ of $\mathbf{H}_k$ contains non-zero elements, i.e., there exist elements $\mathbf{x}_n \in \mathbb{S}^N$, $\mathbf{x}_n \neq \mathbf{0}$, such that $\mathbf{H}_k \mathbf{x}_n = \mathbf{0}$ \citep[ch. 4]{Meyer2001}. As a result, given an observation inversion $\mathbf{x}^{o}_k$, $\mathbf{x}^{o}_k + \mathbf{x}_n$ is also a solution of the equation $\mathbf{H}_k \mathbf{x} = \mathbf{y}^{o}_k$ for any $\mathbf{x}_n \in \mathbb{S}^N$. Therefore, which solution one should take is an open problem in practice. In the context of state estimation, it is desirable to choose a solution that is close to the truth $\mathbf{x}^{tr}_k$, which, unfortunately, is infeasible without the knowledge of $\mathbf{x}^{tr}_k$. As a trade-off, one may choose as a solution some estimate that possesses certain properties. The Moore-Penrose generalized inverse $\mathbf{x}^{o}_k$ given in Eq. (\ref{eq:pseudo_inversion}) is such a choice, which is the unique, and ``best-approximate'', solution in the sense that it has the minimum 2-norm among all least-squares solutions \cite[Theorem 2.5]{Engl2000-regularization}.

It is also worth mentioning what may happen if our assumptions, that $m_y \leq m_x$ and that $\mathbf{H}_k$ is of full row rank, are not valid. In the former case, with $m_y > m_x$, the equation $\mathbf{H}_k \mathbf{x} = \mathbf{y}^{o}_k$ is over-determined, meaning that there may be no solution that solves the equation exactly. One may still obtain an approximate solution by recasting the problem of solving the linear equation as a linear least-squares problem, which yields the unique, least-squares solution in the form of $\mathbf{x}^{o}_k = (\mathbf{H}_k^T \mathbf{H}_k)^{-1}  \mathbf{H}_k^T \mathbf{y}^{o}_k$, similar to (but different from) Eq. (\ref{eq:pseudo_inversion}). Because $\mathbf{H}_k \mathbf{x}^{o}_k - \mathbf{y}^{o}_k$ may not be $\mathbf{0}$ in general, one may thus not be able to find a new estimate $\tilde{\mathbf{x}}^a_k$ with a sufficiently small (e.g., zero) residual. Therefore, the inequality $\Vert \tilde{\mathbf{r}}_k^a \Vert_2 \leq \beta \sqrt{\text{trace}(\mathbf{R}_k)}$ may not hold for some sufficiently small $\beta$. This restriction is consistent with the nature of over-determined problems (that is, no exact solution). It does not necessarily mean that residual nudging cannot be applied to an over-determined problem, but instead implies that the noise level coefficient $\beta$ should entail a lower bound that may be larger than $0$.

In the latter case, without loss of generality, suppose that $m_y \leq m_x$ and $\mathbf{H}_k$ is not of full row rank, then the matrix product $\mathbf{H}_k \mathbf{H}_k^T$ is singular, so that it may be numerically unstable to compute its inverse. In such circumstances, one needs to employ a certain regularization technique to obtain an approximate, but stable, solution. For instance, one may adopt the Tikhonov regularization \citep[ch. 4]{Engl2000-regularization} so that the solution in Eq. (\ref{eq:pseudo_inversion}) becomes $\mathbf{x}^{o}_k = \mathbf{H}_k^T ( \mathbf{H}_k \mathbf{H}_k^T + \alpha \mathbf{I})^{-1} \mathbf{y}^{o}_k$, where $\alpha$ is the regularization parameter chosen according to a certain criterion. The observation inversion in Eq. (\ref{eq:pseudo_inversion}) can be treated as a special case of the Tikhonov regularization solution with $\alpha = 0$, while the concept of residual nudging is also applicable to the general cases with $\alpha \neq 0$ following our deduction in \S~\ref{sec:RN} \footnote{In general cases with $\alpha \neq 0$, it can be shown that a sufficient condition to achieve residual nudging is, for example, $c_k ( \Vert \hat{\mathbf{r}}^a_k \Vert_2 - \Vert \mathbf{y}^o_k \Vert_2) \leq \beta \sqrt{\text{trace}(\mathbf{R}_k)} - \Vert \mathbf{y}^o_k \Vert_2$, with the (possibly) new estimate $\tilde{\mathbf{x}}^a_k$ again given by Eq. (\ref{eq:replacement_mixing}).}. In this sense, the state estimate of the EAKF-RN can be considered as a hybrid of the original EAKF estimate and the (regularized) least-squares solution of the equation $\mathbf{H}_k \mathbf{x} = \mathbf{y}^{o}_k$. This point of view opens up many other possibilities, given the various types of regularization techniques in the literature (see, for example, \citealt{Engl2000-regularization}). 

The computation of the matrix product $\mathbf{H}_k^T ( \mathbf{H}_k \mathbf{H}_k^T)^{-1}$ is a non-trivial issue in large-scale problems, and is worthy of further discussion\footnote{For the experiments to be presented later, since the dimensions of the dynamical models are relatively low, we choose to directly compute the matrix product $\mathbf{H}_k^T ( \mathbf{H}_k \mathbf{H}_k^T)^{-1}$. The matrix inversion $( \mathbf{H}_k \mathbf{H}_k^T)^{-1}$ is done through the MATLAB (R2011b) built-in function INV.}. In general cases where the observation operator $\mathbf{H}_k$ is time varying, the computational cost is comparable to that in evaluating the Kalman gain. In terms of numerical computations, one possible choice is to apply QR factorization \citep[ch. 5]{Meyer2001} to $\mathbf{H}_k^T$ such that $\mathbf{H}_k^T$ is factorized as the product of an orthogonal, $m_x \times m_x$ matrix $\mathbf{Q}$ and an upper-triangular, $m_x \times m_y$ matrix $\mathbf{U}$, where for notational convenience we drop the time index $k$ in these matrices. Note that $\mathbf{Q} \mathbf{Q}^T = \mathbf{Q}^T \mathbf{Q} = \mathbf{I}_{m_x}$, and $\mathbf{U} = [\mathbf{U}_{m_y}^T, \mathbf{0}_{(m_x - m_y)m_y}^T]^T$, with $\mathbf{I}_{m_x}$ being the $m_x$-dimensional identity matrix, $\mathbf{0}_{(m_x - m_y)m_y}$ the $(m_x - m_y) \times m_y$ zero matrix, and $\mathbf{U}_{m_y}$ a non-singular, upper-triangular, $m_y \times m_y$ matrix in which all elements below the main diagonal are zero. With some algebra, it can be shown that the product $\mathbf{H}_k^T ( \mathbf{H}_k \mathbf{H}_k^T)^{-1} =\mathbf{Q} \; [\mathbf{U}_{m_y}^{-1}, \mathbf{0}_{(m_x - m_y)m_y}]^T = \mathbf{Q}_{m_x \, m_y} (\mathbf{U}_{m_y}^{-1})^T$, where $\mathbf{Q}_{m_x \, m_y}$ is a matrix that is comprised of the first $m_y$ columns of $\mathbf{Q}$, and the inverse $\mathbf{U}_{m_y}^{-1}$ of the upper-triangular matrix $\mathbf{U}_{m_y}$ can be computed element-by-element in a recursive way (called back substitution, \citealt[ch. 5]{Meyer2001}). In certain circumstances, further reduction of computational cost and/or storage can be achieved, for instance, when $\mathbf{H}_k$ is sparse \citep[ch. 5]{Meyer2001}; or when $\mathbf{H}_k$ is time invariant, e.g., in a static observation network. In the latter case, one only needs to evaluate the product $\mathbf{H}_k^T ( \mathbf{H}_k \mathbf{H}_k^T)^{-1}$ once and for all.

\section{Numerical results in a linear scalar system} \label{sec:numerical_example_KF}

Here we use a scalar, first order autoregressive (AR1) model driven by Gaussian white noise, to investigate the performance of the Kalman filter (KF,\citealp{Kalman-new}), and that of the KF with residual nudging (KF-RN), in which residual nudging is introduced to the posterior estimate of the KF in the same way as in the EAKF. The motivation in conducting this experiment is the following. With linear and Gaussian observations, the KF provides the optimal estimate in the sense of, for instance, minimum variance \citep{Jazwinski1970}. Therefore, we use the KF estimate as the reference to examine the behaviour of the KF-RN under different settings, which reveals how residual nudging may affect the performance of the KF.

The scalar AR1 model is given by
\begin{linenomath*}
\begin{equation} \label{eq:ar1}
x_{k+1} = 0.9 \, x_k + u_k \, ,
\end{equation}
\end{linenomath*}
where $u_k$ represents the dynamical noise and follows the Gaussian distribution with zero mean and variance 1, and is thus denoted by $u_k \sim N(u_k:0,1)$. The observation model is described by
\begin{linenomath*}
\begin{equation} \label{eq:ar1_obs}
y_{k} = x_k + v_k \, ,
\end{equation}
\end{linenomath*}
where $v_k \sim N(v_k:0,1)$ is the observation noise, and is uncorrelated with $u_k$.

In the experiment, we integrate the AR1 model forward for $10,000$ steps (integration steps hereafter), with the initial value randomly drawn from the Gaussian distribution $N(0,1)$, and the associated initial prior variance being 1. The true states (truth) $\{ x_k \}_{i=1}^{10000}$ are obtained by drawing samples of dynamical noise from the distribution $N(0,1)$, and adding them to $x_k$ to obtain $x_{k+1}$ at the next integration step, and so on. The synthetic observations $y^o_k$ are obtained by adding to model states $x_{k}$ samples of observation noise from the distribution $N(0,1)$. For convenience of comparison, we generate and store synthetic observations at every integration step. However, we choose to assimilate them for every $S_a$ integration steps, with $S_a \in \{1, 2,4, 8\}$, in order to investigate the impact of $S_a$ on filter performance. In doing so, data assimilation with different $S_a$, or other experiment settings (e.g., the noise level coefficient $\beta$ in the KF-RN), will have identical observations at the same integration steps. For convenience, hereafter we may sometimes use the concept ``assimilation step'', with one assimilation step equal to $S_a$ integration steps. In addition, we may also call $S_a$ the assimilation step when it causes no confusion.

In the KF-RN, we also choose to vary the noise level coefficient $\beta$, with $\beta \in \{0.01,0.05,0.1,0.5, 1, 2,3,4,6,8,10 \}$, in order to investigate its effect on filter performance. To reduce statistical fluctuations, we repeat the experiment 20 times, each time with randomly drawn initial value, samples of dynamical and observation noise (so that the truth and the corresponding observations are produced at random). Except for the introduction of residual nudging, the KF-RN have the same configurations and experiment settings as the KF.

We use the average root mean squared error (average RMSE) to measure the accuracy of a filter estimate. For an $m_x$-dimensional system, the RMSE $e_k$ of an estimate $\hat{\mathbf{x}}_k = [\hat{x}_{k,1}, \dotsb, \hat{x}_{k,m_x}]^T$ with respect to the true state vector $\mathbf{x}_k^{tr} = [x_{k,1}^{tr}, \dotsb, x_{k,m_x}^{tr}]^T$ at time instant $k$ is defined as
\begin{linenomath*}
\begin{equation} \label{eq:def_of_rmse}
e_k = \Vert \hat{\mathbf{x}}_k - \mathbf{x}_k^{tr} \Vert_2 /\sqrt{m_x} \, .
\end{equation}
\end{linenomath*}
The average RMSE $\hat{e}_k$ at time instant $k$ over $M$ repetitions of the same experiment is thus defined as $\hat{e}_k =\sum_{j=1}^{M} e_k^j/M$ ($M=20$ in our setting), where $e_k^j$ denotes the RMSE at time instant $k$ in the $j$th repetition of the experiment. We also define the time mean RMSE $\hat{e}$ as the average of $\hat{e}_k$ over the assimilation time window with $N$ integration steps, i.e., $\hat{e} = \sum_{i=1}^{N} \hat{e}_k/N$ ($N = 10000$ here).

We also use the spread to measure the estimated uncertainty associated with an estimation. To this end, let $\hat{\mathbf{P}}_k$ be the estimated covariance matrix with respect to the estimate $\hat{\mathbf{x}}_k$. Then the spread $s_k$ at time instant $k$ is defined as
\begin{linenomath*}
\begin{equation} \label{eq:def_of_spread}
s_k = \sqrt{\text{trace}(\hat{\mathbf{P}}_k) / m_x} .
\end{equation}
\end{linenomath*}
The average spread $\hat{s}_k$ and the time mean (average) spread $\hat{s}$ are defined in a way similar to their counterparts with respect to the RMSE.

Table \ref{table:KF_rmse_spread} reports the time mean RMSEs and spreads of the KF at different assimilation steps $S_a$. The time mean RMSE of the KF grows as $S_a$ increases, indicating that the performance of the KF deteriorates as the assimilation frequency decreases. The time mean spread of the KF exhibits a similar tendency as $S_a$ increases. However, the time mean spread tends to be larger than the time mean RMSE, indicating that the corresponding variance is over-estimated.

Fig. \ref{fig:KF_KFRN_rmse} shows the time mean RMSEs of the KF-RN (dash-dot lines marked by diamonds), as functions of the noise level coefficient $\beta$, at different assimilation steps $S_a$. Given the different orders of magnitudes of $\beta$, we adopt the logarithmic scale for the x-axes. For comparison, we also plot the time mean RMSEs of the KF (solid lines) at each $S_a$. Since the time mean RMSEs of the KF are independent of the choice of $\beta$, they are horizontal lines in the plots. However, the choice of $\beta$ does influence the performance of the KF-RN. As shown in all of the plots of Fig. \ref{fig:KF_KFRN_rmse}, if one adopts a small $\beta$, say $\beta = 0.01$, for the KF-RN, then the resulting time mean RMSE is higher than that of the KF. This is because such a choice may force the KF-RN to rely excessively on the observations when updating the prior estimates, such that the information contents in the prior estimates are largely ignored. As $\beta$ grows, the time mean RMSE of the KF-RN decreases, and eventually converges to that of the KF when $\beta$ is sufficiently large, say $\beta \geq 3$. These results are consistent with our expectation of the behaviour of a filter equipped with residual nudging, as has been discussed in \S~\ref{sec:RN discussion}.

It is also of interest to gain some insights of the behaviour of the fraction coefficients $c_k$ in the KF-RN with different $\beta$. To this end, Fig. \ref{fig:KF_ts_c} plots two sample time series of $c_k$ in the KF-RN with $\beta = 0.1$ (upper left panel), and $\beta = 1$ (lower left panel), respectively, together with their corresponding histograms (right panels). For convenience of visualization, the assimilation time window is shortened to $1000$ steps (with the observations assimilated for every $4$ steps). At $\beta = 0.1$, $c_k$ tends to be relatively small, with the mean value being $0.4213$ and the median $0.3027$. Among the $250$ $c_k$ values, $210$ of them are less than 1, meaning that residual nudging is effective at those steps. A histogram of $c_k$ is also shown on the upper right panel. There it indicates that $c_k$ distributes like a U-shape, with relatively large proportions of $c_k$ taking values that are less than $0.2$, or equal to $1$. On the other hand, at $\beta = 1$, $c_k$ tends to remain close to $1$, with the mean being $0.9892$ and the median $1$, and only $16$ out of $250$ $c_k$ values are less than $1$. These are also manifested in the histogram on the lower right panel, where one can see that $c_k$ largely concentrate on $1$.

In Table \ref{table:KF_rmse_spread} we report the minimum time mean RMSEs that the KF-RN can achieve by varying the value of $\beta$ at different $S_a$, together with the values of the $\beta$ at which the minima are obtained for specific $S_a$. When $S_a = 1,2$, the minimum time mean RMSEs of the KF-RN, both achieved at $\beta =2$, are (very) slightly lower than the time mean RMSEs of the KF; and the time mean RMSEs of the KF-RN become the same as those of the KF when $\beta \geq 3$. On the other hand, when $S_a = 4,8$, the minimum time mean RMSEs of the KF-RN are identical to the time mean RMSEs of the KF, and are obtained when $\beta \geq 2$. The reason that the KF-RN can have lower time mean RMSEs than the ``optimal'' KF at $S_a = 1,2$ might be the following. The classic filtering theory states that the KF is optimal under the minimum variance (MV) criterion \citep{Jazwinski1970}, that is, taking the mean of the posterior conditional pdf as the state estimate, the KF has the lowest possible expectation of squared estimation error. Note that here the expectation is taken over all possible values of the truth (i.e., by treating the truth as a random variable). Therefore, in principle one has to repeat the same experiment for a sufficiently large number of times (with randomly drawn truth) in order to verify the performance of the filters under the MV criterion. For computational convenience, though, we only repeat the experiment $20$ times. Thus in our opinion the slight out-performance of the KF-RN might be largely attributed to statistical fluctuations.

In Table \ref{table:KF_rmse_spread} we do not present the time mean spreads of the KF-RN because they are in fact identical to those of the KF. This is because in the KF, the forecast and update of the (estimated) covariance matrix of the system state are not influenced by the mean estimate of the system state \citep{Jazwinski1970}. Since residual nudging only changes the estimate of the system state (if necessary) and nothing else, it is expected that the KF and KF-RN share the same covariance matrix. This point, however, is not necessarily true in the context of ensemble filtering in a nonlinear system. For instance, if the dynamical model is nonlinear, then the background covariance at the next assimilation time is affected by the analysis mean at the current time, such that two analysis ensembles with different sample (analysis) means but identical sample (analysis) covariance may result in different sample (background) means and covariances at the next assimilation time.

The above results suggest that it may not be very meaningful to introduce residual nudging to a Bayesian filter that already performs well. In practice, though, due to the existence of various sources of uncertainties \citep{Anderson2007,Luo2011_EnLHF}, a Bayesian filter is often sub-optimal, and is even likely to suffer from divergence \citep{Schlee1967}. In such circumstances, instead of only looking into the accuracy of a filter, it may also be desirable to take the stability of the filter into account. Through the experiments below we show that equipping the EAKF with residual nudging can not only help improve its stability, but also achieve a filter accuracy that is comparable to, sometimes even (much) better than, that of the normal EAKF, especially in the small ensemble scenario.

\section{Numerical results in the 40-dimensional L96 model} \label{sec:numerical_example}
\subsection{Experiment settings}
Here we use the $40$-dimensional Lorenz-96 (L96) model \citep{Lorenz-optimal} as the testbed. The governing equations of the L96 model are given by
\begin{linenomath*}
\begin{equation} \label{eq:LE98}
\frac{dx_i}{dt} = \left( x_{i+1} - x_{i-2} \right) x_{i-1} - x_i + F, \, i=1, \dotsb, 40.
\end{equation}
\end{linenomath*}
The quadratic terms simulate advection, the linear term represents internal dissipation, and $F$ acts as the external forcing term \citep{Lorenz-predictability}. Throughout this work, we choose $F = 8$ unless otherwise stated. For consistency, we define $x_{-1}=x_{39}$, $x_{0}=x_{40}$, and $x_{41}=x_{1}$ in Eq.~(\ref{eq:LE98}), and construct the state vector $\mathbf{x} \equiv [x_1,x_2,\dotsb,x_{40}]^T$.

We use the fourth-order Runge-Kutta method to integrate (and discretize) the system from time $0$ to $75$, with a constant integration step of $0.05$. To avoid the transition effect, we discard the trajectory between $0$ and $25$, and use the rest for data assimilation. The synthetic observation $\mathbf{y}_k$ is obtained by measuring (with observation noise) every $d$ elements of the state vector $\mathbf{x}_k = [x_{k,1},x_{k,2},\dotsb,x_{k,40}]^T$ at time instant $k$, i.e.,
\begin{linenomath*}
\begin{equation}
\mathbf{y}_k = \mathbf{H}^d \mathbf{x}_k + \mathbf{v}_k \, ,
\end{equation}
\end{linenomath*}
where $\mathbf{H}^d$ is a $(J+1) \times 40$ matrix such that $\mathbf{H}^d \mathbf{x}_k = [x_{k,1}, x_{k,1+d},\dotsb,x_{k,1+Jd}]^T$, with $J = \text{floor}(39/d)$ being the largest integer that is less than, or equal to, $39/d$, and $\mathbf{v}_k$ is the observation noise following the Gaussian distribution $N(\mathbf{v}_k:\mathbf{0},\mathbf{I}_{J+1})$, with $\mathbf{I}_{J+1}$ being the $(J+1)$-dimensional identity matrix. The elements $(\mathbf{H}^d)_{pq}$ of the matrix $\mathbf{H}^d$ can be determined as follows.
\begin{linenomath*}
\[
(\mathbf{H}^d)_{pq} = 1~\text{if}~q=(p-1)d+1\, , ~\text{otherwise}~(\mathbf{H}^d)_{pq} = 0 \, ,
\]
\end{linenomath*}
for $p = 1, \dotsb, (J+1), ~ q = 1,\dotsb, 40$. In all the experiments below, we generate and store the synthetic observations at every integration step, but assimilate the observations for every $4$ integration steps unless otherwise stated.

The filters in the experiments are configured as follows. To generate an initial background ensemble, we run the L96 model from $0$ to $2500$ (overall $50000$ integration steps), and compute the temporal mean and covariance of the trajectory\footnote{Let $\{\mathbf{x}_k\}_{k=1}^N$ be a set of state vectors at different time instants which form a state trajectory from time instant $1$ to $N$. Then the temporal mean and covariance of the trajectory are taken as the sample mean and covariance of the set $\{\mathbf{x}_k\}_{k=1}^N$, respectively.}. We then assume that the initial state vectors follow the Gaussian distribution with the same mean and covariance, and draw a specified number of samples to form the background ensemble. Covariance inflation \citep{Anderson-Monte} and localization \citep{Hamill-distance} are conducted in all the experiments. Concretely, covariance inflation, with the inflation factor $\lambda$, is introduced following the discussion in \S~\ref{subsec:filtering_EAKF}. Covariance localization is conducted following \citet{Anderson2007,Anderson2009}, which introduces an additional parameter $l_c$, called the length scale (or half-width following \citealt{Anderson2007,Anderson2009}) hereafter, to the EAKF. The distance $d_{ij}$ between two state variables $x_i$ and $x_j$ are defined as $d_{ij} = \text{min}(|i-j|/40,1-|i-j|/40)$, and the corresponding tapering coefficient $\eta_{ij}$ (cf. the text below Eq.~(\ref{eq:eakf_cross_variance})) is determined by the fifth-order polynomial function $\xi(d_{ij},l_c)$ in \citet{Gaspari1999} with half-width $l_c$. For $d_{ij} < 2 \, l_c$, one has $0 < \eta_{ij} \leq 1$, and $\eta_{ij} = 0$ otherwise. With both covariance inflation and localization, the performance of the normal EAKF is in general comparable to the established results with respect to the L96 model under similar experiment setting, see, for example, \citet{fertig2007comparative,hunt2004four}.

To reduce statistical fluctuations, we repeat each experiment below for $20$ times, each time with randomly drawn initial state vector, initial background ensembles and observations. Except for the introduction of residual nudging, in all experiments the normal EAKF and the EAKF-RN have identical configurations and experiment settings.

\subsection{Experiment results}

\subsubsection{Results with different observation operators}
Here we consider four different observation operators $\mathbf{H}^d$, with $d = 1, 2, 4, 8$, respectively. For convenience, we refer to them as the full, 1/2, 1/4 and 1/8 observation scenarios, respectively. The concrete configurations of the normal EAKF and the EAKF-RN are the following. In both filters the ensemble size is fixed to be $20$. The half-width $l_c$ of covariance localization increases from $0.1$ to $0.5$, each time with an even increment of $0.1$. For convenience we denote this setting by $l_c \in \{0.1:0.1:0.5\}$. Similar notations will be frequently used later. The inflation factor $\lambda \in \{1:0.05:1.25\}$, and the noise level coefficient $\beta = 2$ in the EAKF-RN.

The upper panels of Fig. \ref{fig:normal_and_darn_EAKF_rmse} shows the contour plots of the time mean RMSEs of the normal EAKF (left), and that of the EAKF-RN (right), in the full observation scenario, as functions of the inflation factor $\lambda$ and the half-width $l_c$. Given a fixed $\lambda$, the time mean RMSEs of both the EAKF and EAKF-RN tend to increase as the half-width $l_c$ increases. On the other hand, given a fixed $l_c$, when $l_c = 0.1$, the time mean RMSEs of both filters exhibit the U-turn behaviour, i.e., the time mean RMSEs tend to decrease as $\lambda$ grows, until it reaches a certain value ($1.10$ for both filters). After that, the time mean RMSEs will increase instead as $\lambda$ grows further. However, when $l_c > 0.1$, the time mean RMSEs of both filters tend to decrease as $\lambda$ increases within the range of tested $\lambda$. The normal EAKF achieves its minimum time mean RMSE (0.5605) at the point $(l_c = 0.1, \lambda = 1.10)$, and the EAKF-RN also hits its minimum time mean RMSE (0.5586) at the same place. In general, the EAKF and the EAKF-RN have similar performance at $l_c = 0.1$, but at other places the EAKF-RN may perform substantially better than the EAKF. For instance, at $(l_c = 0.4,\lambda=1.05)$ the time mean RMSE of the normal EAKF is about $3.3$, while that of the EAKF-RN is about $1.6$. Moreover, a filter divergence is spotted in the normal EAKF at $(l_c = 0.3,\lambda=1.25)$, so that the contour plot around this point is empty and indicates no RMSE value. Filter divergence, however, is not observed in the EAKF-RN at the same place. For clarity, here a ``divergence'' is identified as an event in which the RMSE of a filter becomes abnormally large. More specifically, the filter is considered divergent in the Lorenz 96 model, if its RMSE at any particular time instant is larger than $10^3$. As mentioned previously, we repeat each experiment 20 times in order to reduce statistical fluctuations. In accordance with this setting, a filter divergence is reported whenever there is at least one (but not necessarily all) divergence(s) out of 20 repetitions.

In the 1/2 and 1/4 observation scenarios, there are many cases in which filter divergences are spotted. For this reason, we choose to directly report the assimilation results in Tables \ref{table:normal_EAKF_rmse_half} and \ref{table:normal_EAKF_rmse_quarter}, respectively, rather than show their contour plots as in the full observation scenario. In the 1/2 observation scenario, filter divergences of the normal EAKF, marked by ``Div'' in Table \ref{table:normal_EAKF_rmse_half}, are spotted in 24 out of 30 different combinations of $l_c$ and $\lambda$ values (5 $l_c$ values by 6 $\lambda$ values). In contrast, in the EAKF-RN no filter divergence is observed. On the other hand, when there is no filter divergence occurring in either filter, the performance of the EAKF and the EAKF-RN is very close to each other, with the time mean RMSEs of the EAKF-RN slightly lower than those of the EAKF, except at $(l_c = 0.1,\lambda=1.15)$ and $(l_c = 0.1,\lambda=1.25)$. The situation in the 1/4 observation is similar. As shown in Table \ref{table:normal_EAKF_rmse_quarter}, the EAKF diverges in 17 out of 30 tested cases, while there is no filter divergence spotted in the EAKF-RN. The performance of the EAKF and the EAKF-RN is  close to each other when the EAKF does not diverge.

The lower panels of Fig. \ref{fig:normal_and_darn_EAKF_rmse} shows the contour plots of the time mean RMSEs of the normal EAKF (left), and that of the EAKF-RN (right), in the 1/8 observation scenario. In this scenario, no filter divergence is spotted in the EAKF. Overall, the performance of the EAKF and the EAKF-RN is very close to each other, although the EAKF-RN has a slightly lower minimum time mean RMSE (2.9556 achieved at $(l_c = 0.1,\lambda=1)$) than that of the EAKF (2.9619 obtained at the same place).

We then examine the impact of residual nudging on the time mean spreads of the filters in different observation scenarios. For the full and 1/8 observation scenarios, we plot the time mean spreads of the EAKF and the EAKF-RN in Fig.~\ref{fig:normal_and_darn_EAKF_spread}; while for the 1/2 and 1/4 observation scenarios, we report them in Tables \ref{table:normal_EAKF_rmse_half} and \ref{table:normal_EAKF_rmse_quarter}, in the parentheses after the RMSE values. In all the reported cases in which the EAKF does not diverge, the time mean spreads of the EAKF-RN in general do not significantly deviate from those of the EAKF. In cases that the EAKF does diverge, the EAKF-RN may still maintain positive and finite time mean spreads. The closeness of the time mean spreads of the EAKF and EAKF-RN in the former cases, though, may depend on the experiment settings, e.g., the choice of the noise level coefficient $\beta$. However, from our experience, as long as $\beta$ is reasonably large (say $\beta \geq 2$), the time mean spread of the EAKF-RN often approaches that of the EAKF. For brevity, hereafter we do not report the spread values any more.

Overall, in both the normal EAKF and the EAKF-RN, their time mean RMSEs tend to increase as the number of elements in an observation decreases. The performance of the EAKF-RN, in terms of time mean RMSE, is in general comparable to, and sometimes (substantially) better than, that of the EAKF. Moreover, the EAKF-RN tends to perform more stably than the EAKF.

\subsubsection{Results with different noise level coefficients} \label{subsubsec:noise_level}
Next we examine the effect of the noise level coefficient $\beta$ on the performance of the EAKF-RN. The experiment settings are as follows. We conduct the experiments in four observation scenarios as in the previous experiment. The ensemble size of the EAKF-RN is $20$. We choose the noise level coefficient $\beta$ from the sets $\{0\}$, $\{ 0.02:0.02:0.1 \}$, $\{ 0.2:0.2:1 \}$, and $\{2, 3, 4, 6, 8\}$. The reason to single out $\beta=0$ will be given soon. Under the above setting, it is infeasible for us to adopt too many combinations of $l_c$ and $\lambda$ as in the previous experiment, either for presentation or computation. Therefore, we only choose two such combinations in the current experiment (similar choices will also be made in subsequent experiments, in which we can only afford to vary some of the parameter values, and have to freeze the rest). In the first combination we let $l_c = 0.1$ and $\lambda = 1.15$, and in the second $l_c = 0.3$ and $\lambda = 1.05$. From the previous experiment results, the former choice represents a relatively good filter configuration for the normal EAKF, while the latter a less proper one. We thus use these two configurations to illustrate the effect of residual nudging when the normal EAKF has reasonable/(relatively) poor performance.

Fig.~\ref{fig:EAKF_rmse_vs_noise_level_obsOP_lc01_delta015} depicts the time mean RMSEs of the EAKF-RN as functions of $\beta$ in different observation scenarios, in which the relatively good filter configuration $l_c = 0.1$ and $\lambda = 1.15$ is adopted. Due to different orders of magnitudes of $\beta$, the x-axes are all plotted in the logarithmic scale. For this reason, it is inconvenient to show the results of $\beta = 0$ at $\log 0$ $(= - \infty)$. Instead, we plot the results at $\beta = 0.005$, and ``artificially'' label that point 0. The time mean RMSEs of the normal EAKF are independent of $\beta$, and are plotted as horizontal lines in the relevant sub-figures (if no filter divergence in the normal EAKF). In all observation scenarios, the time mean RMSEs of the EAKF-RN are relatively large at small $\beta$ values (say $\beta = 0.02$). As $\beta$ increases, the time mean RMSEs of the EAKF-RN tend to converge to those of the normal EAKF. During the processes of convergence, the minimum time mean RMSE of the EAKF-RN in the full observation scenario is lower than that of the normal EAKF, while the minimum time mean RMSEs of the EAKF-RN in other observation scenarios are either indistinguishable from (in the 1/2 and 1/4 observation scenarios), or slightly higher than (in the 1/8 observation scenario), those of the normal EAKF.

Fig.~\ref{fig:EAKF_rmse_vs_noise_level_obsOP_lc03_delta005} shows the time mean RMSEs of the normal EAKF and the EAKF-RN, with experiment settings similar to those in Fig.~\ref{fig:EAKF_rmse_vs_noise_level_obsOP_lc01_delta015}, except that the covariance localization and inflation configuration becomes $l_c = 0.3$ and $\lambda = 1.05$, respectively, which, as will be shown below, makes the normal EAKF perform worse in comparison to the previous case in Fig.~\ref{fig:EAKF_rmse_vs_noise_level_obsOP_lc01_delta015}.

With $l_c = 0.3$ and $\lambda = 1.05$, the resulting EAKF-RN behaves similarly to that with the previous configuration $l_c = 0.1$ and $\lambda = 1.15$. For the current filter configuration, though, as $\beta$ grows, the time mean RMSEs of the EAKF-RN exhibit clear troughs in all observation scenarios. On the other hand, compared to the previous results in Fig.~\ref{fig:EAKF_rmse_vs_noise_level_obsOP_lc01_delta015}, the performance of the normal EAKF deteriorates in all observation scenarios. Indeed, with the current filter configuration, the normal EAKF may perform (substantially) worse than the EAKF-RN under the same experiment settings, especially if a proper $\beta$ value is chosen for the EAKF-RN. In particular, the normal EAKF diverges in the 1/2 (upper right) and 1/4 (lower left) observation scenarios, while no filter divergence is spotted in the EAKF-RN with $\beta \leq 3$, although the EAKF-RN does diverge in the 1/2 and 1/4 observation scenarios, given $\beta \geq 4$. This suggests that one may increase the stability of the EAKF-RN against filter divergence by decreasing the value of $\beta$, so that $c_k$ is closer to 0 and the observation inversion becomes more influential in Eq.~(\ref{eq:replacement_mixing}), as we have discussed in \S\ref{sec:RN discussion}.

It is also worth mentioning the behaviour of the EAKF-RN with small $\beta$ values. As one can see in Figs. \ref{fig:EAKF_rmse_vs_noise_level_obsOP_lc01_delta015} and \ref{fig:EAKF_rmse_vs_noise_level_obsOP_lc03_delta005}, given different filter configurations, the EAKF-RN may behave quite differently at relatively large $\beta$ values. However, as $\beta$ tends to 0, the time mean RMSEs of the EAKF-RN with different configurations tend to converge, despite the different combinations of $l_c$ and $\lambda$. This is because, as $\beta \rightarrow 0$, $c_k \rightarrow 0$ in Eq. (\ref{eq:fraction_coefficient}), hence the new estimate $\tilde{\mathbf{x}}^a_k$, according to Eq. (\ref{eq:replacement_mixing}), approaches the observation inversion $\mathbf{x}^{o}_k$, which is independent of, for instance, the half-width $l_c$, the inflation factor $\lambda$ and the ensemble size\footnote{When the observation operator is time-varying, the assimilation step $S_a$ in general has an influence on the observation inversion, as $S_a$ decides when the observations are assimilated.}. Since the time mean RMSE continuously depends on $\beta$, it is not surprising to find that in Figs. \ref{fig:EAKF_rmse_vs_noise_level_obsOP_lc01_delta015} and \ref{fig:EAKF_rmse_vs_noise_level_obsOP_lc03_delta005}, the time mean RMSEs of the EAKF-RN with small $\beta$, say at $\beta = 0.02$, are very close to the corresponding values at $\beta = 0$.

More insights of the filters' behaviour may be gained by examining the fraction coefficient $c_k$ in the EAKF-RN. For the relatively good filter configuration ($l_c = 0.1$ and $\lambda = 1.15$), we have seen in Fig. \ref{fig:EAKF_rmse_vs_noise_level_obsOP_lc01_delta015} that the EAKF and the EAKF-RN have very close performance, and our experiment results show that $c_k$ mostly concentrate on $1$, similar to the situations on the lower panels of Fig. \ref{fig:KF_ts_c} (not reported). Of more interest is the case in which the normal EAKF is less properly configured ($l_c = 0.3$ and $\lambda = 1.05$), and may suffer from filter divergence. On the upper panels of Fig. \ref{fig:EAKF_ts_c} we show sample time series of the RMSEs of the normal EAKF and EAKF-RN ($\beta = 2$) in the 1/2 observation scenario. On the upper left panel, the EAKF has an exceptionally large RMSE (in the order of $10^{21}$) at time step $k = 26$, is thus considered diverged. In contrast, on the upper right panel, the EAKF-RN ($\beta = 2$) has all the RMSEs less than $5$ (with the corresponding time mean RMSE being $1.8931$), and filter divergence is avoided. The lower left panel shows the time series of the fraction coefficient $c_k$, which has the mean $0.9499$ and the median $1$. Among $250$ $c_k$ values, $78$ are less than $1$. 
For reference, a histogram of $c_k$ is plotted on the lower right panel, which confirms that $c_k$ largely concentrate on $1$.

In Fig. \ref{fig:EAKF_ts_c_interval} we also examine what happens before the normal EAKF diverges. On the upper panel, we show the time series of the RMSEs of the EAKF (in the solid line with asterisks) and the EAKF-RN ($\beta = 2$, in the dotted line with plus signs). One can see that, at the beginning, say, when the time instant $k \leq 15$, the difference between the EAKF and the EAKF-RN is relatively less significant. For $16 \leq k \leq 25$, the difference becomes more obvious. On the middle panel we report the difference between the EAKF and the EAKF-RN ($\beta = 2$), in terms of the RMSE of the EAKF minus that of the EAKF-RN, for $1 \leq k \leq 16$. The reason for not including the RMSE differences at larger time instants is that their amplitudes are relatively large and may make relatively small values indistinguishable from $0$, which is not desired for our purpose. On the lower panel, we also show the fraction coefficients $c_k$ of the EAKF-RN ($\beta = 2$) for $1 \leq k \leq 25$. Note the availability of $c_k$ depends on the availability of the incoming observations, therefore $c_k$ appear for every $4$ steps only. Based on these figures, one may tell what happens to make the EAKF and EAKF-RN behave differently. At time step $k=4$, there is an incoming observation. However, because $c_4 = 1$, the EAKF and EAKF-RN share identical estimates from $k=1$ to $k=7$. At $k=8$, there is one more incoming observation, and this time $c_8$ is less than $1$, meaning that residual nudging is effective, so that there is a (very) small difference spotted between the estimates of the EAKF and EAKF-RN. At $k=12$, residual nudging is conducted again (but no more for subsequent steps up to $k=24$), which, together with the previous residual nudging, makes the estimates of the EAKF-RN deviate from those of the EAKF, and eventually avoid filter divergence at $k=26$.

Overall, we have shown that, when the normal EAKF is properly configured, the performance of the normal EAKF and the EAKF-RN is in general comparable. However, if the EAKF is not configured properly, then the EAKF-RN may perform (substantially) better than the normal EAKF. For many large scale data assimilation problems, it may be very expensive to conduct an extensive parameter searching in order to configure the EnKF \citep{Anderson2007}. Should the EnKF be ill-configured, we expect that introducing residual nudging to the EnKF may enhance its performance, in terms of filter accuracy and/or stability against divergence.

\subsubsection{Results with different ensemble sizes}
Here we examine the effect of the ensemble size $n$ on the performance of the normal EAKF and the EAKF-RN. The experiment settings are as follows. We also conduct the experiment in four observation scenarios. The ensemble size $n$ is chosen from the set $\{2, 4, 6, 8, 10, 20, 40, 60, 80 \}$. In the experiment we fix $l_c=0.1$ and $\lambda = 1.15$ for both the normal EAKF and the EAKF-RN. In the EAKF-RN, we adopt two noise level coefficients, with $\beta$ being 1 and 2, respectively.

Fig.~\ref{fig:EAKF_RN_rmse_vs_ensize} shows the time mean RMSEs of the normal EAKF (solid lines with squares), and those of the EAKF-RNs with  $\beta = 1$ and $2$ (dotted lines with bold points, and dash-dotted lines with crosses, respectively), in different observation scenarios. In the full observation scenario, no filter divergence is found for all the ensemble sizes $n$ in either filter. When $n \leq 10$, the EAKF-RN with $\beta =1$ tends to perform better than the EAKF-RN with $\beta =2$, while the latter is better than the normal EAKF. This is particularly the case with a relatively small ensemble size, say at $n=2$. On the other hand, when $n \geq 20$, the time mean RMSEs of the three filters are almost indistinguishable.

In the 1/2 observation scenario, the normal EAKF diverges when $n \leq 10$, so there are no square markers appearing at those $n$ values. The EAKF-RN with $\beta = 2$ appears more robust than the normal EAKF, although there is still a filter divergence spotted at $n=4$. In contrast, the EAKF-RN with $\beta=1$ is the most robust filter, which does not diverge for all the tested ensemble sizes. In terms of time mean RMSE, though, when the filters do not diverge, the EAKF-RN with $\beta = 1$ tends to perform worse than the EAKF-RN with $\beta=2$, while the latter appears to be indistinguishable from the normal EAKF for $n \geq 20$.

The situations in the 1/4 and 1/8 observation scenarios are similar to that in the 1/2 one. In the 1/4 observation scenario, the normal EAKF diverges for $n \leq 8$, while the EAKF-RN appears to be more robust, except that there is a filter divergence at $n =4$ for the EAKF-RN with $\beta =2$. When $n=2$, the EAKF-RN with $\beta =2$ performs better than the filter with $\beta =1$, but at $n=6$ or $8$, the filter with $\beta =1$ performs better instead. For $n \geq 10$, the performance of all three filters are almost indistinguishable. In the 1/8 observation scenario, the normal EAKF and the EAKF-RN with $\beta=2$ diverge at $n=2$ and $4$, while the EAKF-RN with $\beta = 1$ diverges only at $n=2$. For $n=6$ or $8$, the EAKF-RN with $\beta=1$ has the best performance in terms of time mean RMSE, the EAKF-RN with $\beta =2$ the second, while the normal EAKF the last. For $n \geq 10$, the performance of the three filters are almost indistinguishable, except that at $n=10$, the time mean RMSE of the EAKF-RN with $\beta=1$ is slightly higher than those of the other two filters.

The above results suggest that $n=20$ appears to be a reasonable ensemble size for the normal EAKF in the L96 model, since in all these four observation scenarios, the performance of the normal EAKF with $n=20$ is very close to that with larger $n$ values. As the ensemble size $n$ decreases, the normal EAKF becomes more unstable. The performance of the EAKF-RN with $\beta=1$ and $2$ is almost indistinguishable from the normal EAKF for $n \geq 20$. However, given smaller ensemble sizes, the EAKF-RN tends to perform better than the normal EAKF, in terms of both filter accuracy and stability against filter divergence. In particular, one may enhance the stability of the EAKF-RN by reducing the noise level coefficient $\beta$, since as $\beta \rightarrow 0$, the time mean RMSEs of the EAKF-RN in different observation scenarios become independent of the ensemble size $n$, and approach the corresponding values at $\beta = 0$. This property may be of interest in certain circumstances, for instance, those in which, due to practical limitations, one can only afford to run an EnKF with a very small ensemble size, so that filter stability becomes an important factor in consideration.

\subsubsection{Results with different assimilation steps and observation noise variances}
Here we examine the effects of the assimilation step $S_a$ and the observation noise variance on the performance of the normal EAKF and the EAKF-RN. We assume that the observation noise covariance matrix $\mathbf{R}_k$ is in the form of $\gamma \mathbf{I}$, where $\mathbf{I}$ is the identity matrix with a suitable dimension in different observation scenarios, and $\gamma >0$ is a real scalar. As a result, the variances of $\mathbf{R}_k$ are $\gamma$ for all variables in an observation vector, while the cross-variances are all zero. The experiment settings are the following. The ensemble size is $20$, $l_c=0.1$ and $\lambda = 1.15$ for both the normal EAKF and the EAKF-RN. The noise level coefficients $\beta$ is 2 in the EAKF-RN. We conduct the experiment in four different observation scenarios, and choose $S_a$ from the set $\{1,4,8,12\}$, and $\gamma$ from the set $\{0.01,0.1,1,10,50\}$. The relatively large values of $\gamma$, say $\gamma = 10,50$, are used to represent the scenario in which the quality of the observations is relatively poor. Here we assume that we know the observation noise variance precisely, while in a subsequent experiment we will consider the case in which the observation noise variance is mis-specified.

Figs. \ref{fig:normal_rmse_varying_amStep_obsLvl} and \ref{fig:darn_rmse_varying_amStep_obsLvl} show the time mean RMSEs of the normal EAKF and the EAKF-RN, respectively, in different observation scenarios. In the full observation scenario (upper left panels), for a fixed variance $\gamma$, the time mean RMSEs of both the normal EAKF and the EAKF-RN tend to increase as the assimilation step $S_a$ increases. On the other hand, for a fixed $S_a$, the time mean RMSEs of both filters appear to be monotonically increasing functions of the variance $\gamma$. With $\gamma =0.01,0.1, 1$, the time mean RMSEs of the EAKF-RN tend to be lower than those of the normal EAKF, while with $\gamma = 10, 50$, they are almost indistinguishable, meaning that for relatively poor observation, the normal EAKF and the EAKF-RN have almost the same performance in terms of estimation accuracy, which appears to be also true in other observation scenarios, as will be shown below. In terms of filter stability, for $S_a = 8$ and $12$, the normal EAKF diverges at $\gamma=0.01$ and $0.1$, but the EAKF-RN avoids filter divergences at all these places.

In the 1/2 observation scenario (upper right panels), for a fixed variance $\gamma$, the time mean RMSEs of both the normal EAKF and the EAKF-RN also grow as the assimilation step $S_a$ increases. However, for a fixed $S_a$, the time mean RMSEs of the two filters have behaviour different from that in the previous observation scenario. For $S_a =1$, the time mean RMSE of the normal EAKF is still a monotonically increasing function of $\gamma$; for $S_a = 4,8$, the normal EAKF diverges at $\gamma=0.01$ and $0.1$, and has monotonically increasing time mean RMSE for $\gamma \geq 1$; for $S_a = 12$, the time mean RMSE of the normal EAKF achieves its minimum at $\gamma=0.1$ (slightly lower than that at $0.01$), and thus exhibits the U-turn behaviour, a phenomenon that is more visible in the EAKF-RN. Indeed, for all tested $S_a$ values, the time mean RMSEs of the EAKF-RN all have their minima at $\gamma=0.1$, rather than at $\gamma = 0.01$.
The normal EAKF and the EAKF-RN have almost indistinguishable time mean RMSEs for $\gamma \geq 1$. While the normal EAKF tends to perform better than the EAKF-RN at $\gamma=0.01$ and $0.1$ in terms of time mean RMSE, it is more likely to suffer from filter divergence (e.g., at $S_a = 4,8$). This is an example of the trade-off between filter accuracy and stability, as discussed in \S \ref{sec:RN discussion}.

In the 1/4 observation scenario (lower left panels), for a fixed assimilation step $S_a$, the time mean RMSEs of both the normal EAKF and the EAKF-RN again appear to be monotonically increasing as $\gamma$ increases. For a fixed variance $\gamma$, though, the time mean RMSEs of both filters tend to exhibit the U-turn behaviour, in which the minimum time mean RMSE is achieved at $S_a =4$ (except for the filter divergence in the normal EAKF at $\gamma=0.01$), rather than at $S_a =1$. The normal EAKF and the EAKF-RN have almost indistinguishable time mean RMSEs for $\gamma \geq 0.1$. At $\gamma = 0.01$, though, the normal EAKF seems to perform better than the EAKF-RN in terms of time mean RMSE. However, filter divergences are spotted at $(S_a=4,\gamma=0.01)$ and $(S_a=1,\gamma=50)$, which are again avoided in the EAKF-RN.

In the 1/8 observation scenario (lower right panels), the quantitative behaviour of the two filters, as functions of $S_a$ and $\gamma$, is almost the same as that in the 1/4 observation scenario. The main differences are the following. The time mean RMSEs of the normal EAKF and the EAKF-RN are almost indistinguishable in all tested cases. Filter divergences are spotted at $S_a=1$, with $\gamma=1$, $10$ and $50$, respectively, not only in the normal EAKF, but also in the EAKF-RN. One may, however, avoid these filter divergences in the EAKF-RN by assigning to it a smaller $\beta$, as some of the previous experiment results have suggested.

Overall, the above experiment results are consistent with our discussion in \S~\ref{sec:RN discussion}. When equipped with residual nudging, the EAKF-RN appears to be more stable than the normal EAKF, although maybe at the cost of some loss of estimation accuracy in certain circumstances (e.g., when with too small $\beta$ values).

\subsubsection{Results with imperfect models and mis-specified observation error covariances}
Finally, we examine filter performance of the normal EAKF and the EAKF-RN when they are subject to uncertainties in specifying the forcing term $F$ in Eq.~(\ref{eq:LE98}) and the observation error covariance $\mathbf{R}_k$. We again conduct the experiments in four observation scenarios. The ensemble sizes of both filters are $20$. The half-width $l_c$ of covariance localization is $0.1$, and the covariance inflation factor $\lambda$ is $1.15$. The true value of $F$ is 8, while the true observation error covariance $\mathbf{R}_k$ is $\mathbf{I}_{20}$. In the experiments we let the value of $F$ in the (possibly) imperfect model be chosen from the set $\{ 4, 6, 8, 10, 12 \}$, and the (possibly) mis-specified covariance $\mathbf{R}_k$ in the form of $\gamma \mathbf{I}_{20}$, with $\gamma \in \{ 0.25, 0.5, 1, 2, 5, 10\}$ \footnote{The (possibly) mis-specified observation error covariance, in the form of $\gamma \mathbf{I}_{20}$, is used for both background update, as described in \S\ref{subsec:filtering_EAKF}, and residual nudging through Eq.~(\ref{eq:fraction_coefficient}).}. In the EAKF-RN the noise level coefficient $\beta = 2$.

Figs. \ref{fig:normal_rmse_varying_F_and_c} and \ref{fig:darn_rmse_varying_F_and_c} show the time mean RMSEs of the normal EAKF and the EAKF-RN, respectively, as functions of the (possibly) mis-specified driving force $F$ and the observation noise variance $\gamma$, in different observation scenarios. In the full observation scenario (upper left panels), for a fixed $\gamma$, the time mean RMSEs of both filters exhibit the U-turn behaviour with respect to $F$, achieving their minima at $F=8$. This point also appears to be valid in other observation scenarios. On the other hand, for a fixed $F$, the behaviour of the filters is very similar to that reported in Figs. \ref{fig:EAKF_rmse_vs_noise_level_obsOP_lc01_delta015} and \ref{fig:EAKF_rmse_vs_noise_level_obsOP_lc03_delta005}, since the role of the (possibly) mis-specified variance $\gamma$ is similar to the observation noise level coefficient $\beta$ (note, though, that $\gamma$ also appears in the computation of the Kalman gain). When $\gamma$ is relatively small (say $\gamma \leq 2$), the EAKF-RN tends to perform better than the normal EAKF in terms of time mean RMSE. Moreover, the normal EAKF diverges at $(F=12,\gamma=0.25)$, while the EAKF-RN avoids the divergence. On the other hand, when $\gamma$ is relatively large (say $\gamma \geq 6$), the EAKF-RN and the normal EAKF have almost indistinguishable performance, not only for the current experiment results, but also for those in the other observation scenarios. This is largely because mistakenly over-estimating the variance $\gamma$ has an effect similar to increasing $\beta$, so that the observation inversion in Eq. (\ref{eq:replacement_mixing}) becomes less influential for state estimation, and the EAKF-RN has almost the same estimate as the normal EAKF.

In the 1/2 observation scenario (upper right panels), when $\gamma$ is relatively small (say $\gamma \leq 1$), the normal EAKF tends to diverge for all $F$. The EAKF-RN avoids filter divergences in some of the areas, though there are still two cases spotted at $F=12$, with $\gamma=1$ and $\gamma=2$, respectively. As $\gamma$ becomes larger, the performance of the normal EAKF and the EAKF-RN are very close to each other, similar to the situation in the full observation scenario. In both the 1/4 and 1/8 observation scenarios (lower panels), there are also almost no differences between the time mean RMSEs of the two filters, although the time mean RMSE of the EAKF-RN appears to be slightly lower than that of the normal EAKF in the 1/4 observation scenario for relatively small $F$ and $\gamma$ (around the lower left corners). Both filters diverge in the 1/4 observation scenario, at $(F=10,\gamma=0.25)$, otherwise neither filter diverges.

\section{Discussion and conclusion} \label{sec:conclusion}
In this work we proposed an auxiliary technique, called residual nudging, for ensemble Kalman filtering. The main idea of residual nudging is to monitor, and if necessary, adjust the residual norm of a state estimate. In an under-determined state estimation problem, if the residual norm is larger than a pre-specified value, then we reject the estimate and replace it by a new one whose residual norm is equal to the pre-specified value; otherwise we accept the estimate. We discussed how to choose the pre-specified value, and demonstrated how one can construct a new state estimate based on the original one and the observation inversion, given a linear observation operator.

Through the numerical experiments in both the scalar AR1 and the Lorenz 96 models, we showed that, by choosing a proper noise level coefficient, the ensemble adjustment Kalman filter with residual nudging (EAKF-RN) in general works more stably than the normal EAKF, while achieving an accuracy that is often comparable to, sometime even (much) better than that of the normal EAKF, especially if the normal EAKF is ill-configured. This may occur, for instance, when the EAKF is equipped with improperly chosen covariance inflation factor and/or half-width of covariance localization, too small ensemble size, and so on. In many data assimilation practices, it may be very expensive to conduct extensive searching for proper inflation factor and/or half-width, or to run a large scale model with too many ensemble members. In such circumstances, we expect that residual nudging may help improve the filter performance, in terms of filter stability, and even accuracy. 

We also implemented residual nudging in some other filters, including the stochastic ensemble Kalman filter \citep{Burgers-analysis} and the singular evolutive interpolated Kalman filter (SEIK) \citep{Hoteit2002,Pham2001}, and observed similar performance improvements (not shown in this work). Since residual nudging only aims to adjust the estimates, we envision that residual nudging can be associated with other data assimilation approaches, including, for instance, the extended Kalman filter, the particle filter, and various smoothers. This will be verified elsewhere.

One problem not addressed in this work is the nonlinearity of the observation operator. In such circumstances, we conjecture that the rule in choosing the pre-specified value $\beta \sqrt{\text{trace}(\mathbf{R}_k)}$ may still be applicable. However, the construction of new state estimates would become more complicated than Eqs.~(\ref{eq:fraction_coefficient}) and (\ref{eq:replacement}). One possible strategy is to linearize the observation operator, or employ more sophisticated methods, such as iterative searching algorithms (see, for example, \citealt{Gu2007-iterative,Lorentzen2011-iterative}), to find new estimates whose residual norms are no larger than $\beta \sqrt{\text{trace}(\mathbf{R}_k)}$. This is another topic that will be investigated in the future.

\section*{Acknowledgement}

We thank Dr Jeffrey Anderson for his kind advices on using the EAKF codes (MATLAB) in the Data Assimilation Research Testbed (DART, version "kodiak", 2011). This provides the basis for us to build the EAKF codes used in our experiments.

We also thank two anonymous reviewers for their constructive comments and suggestions. One reviewer points out the similarity between residual nudging and the adaptive inflation schemes in \citet{Anderson2007,Anderson2009}, and suggests conducting the experiment with respect to the KF in the AR1 model. Another reviewer points out the possible combinations of the EnKF and the regularization approaches in inverse problems.

Luo acknowledges partial financial support from the Research Council of Norway and industrial partners, through the project "Transient well flow modelling and modern estimation techniques for accurate production allocation".

\bibliographystyle{tellus}
\bibliography{references}

\renewcommand{\thefigure}{\arabic{figure}}
\renewcommand{\thetable}{\arabic{table}}
\clearpage

\listoftables
\listoffigures

\clearpage
\begin{table*} 
\centering
\caption{\label{table:KF_rmse_spread} Time mean RMSEs and spreads of the KF, and the minimum time mean RMSEs (over different $\beta$) of the KF-RN, in the AR1 model with different $S_a$. The KF and KF-RN have identical time mean spreads, therefore only those of the KF are presented. In the bottom row we also report the ranges of $\beta$ in which the minimum time mean RMSEs of the KF-RN are achieved.}
\begin{tabular}{ccccc}
\hline \hline
\multirow{2}{*}{KF}& \multicolumn{4}{c}{$S_a = $} \\
\cline{2-5}
& 1 & 2 & 4 & 8 \\
\hline
RMSE & 0.6184 & 0.8260 & 1.0592 & 1.2997 \\
spread & 0.7729 & 1.0413 & 1.3419 & 1.8241 \\
\hline \hline
\multirow{2}{*}{KF-RN}& \multicolumn{4}{c}{$S_a = $} \\
\cline{2-5}
& 1 & 2 & 4 & 8 \\
\hline
min RMSE & 0.6183 & 0.8259 & 1.0592 & 1.2997 \\
achieved at & $\beta = 2$ & $\beta = 2$ & $\beta \geq 2$ & $\beta \geq 2$ \\
\hline \hline
\end{tabular}
\end{table*}

\clearpage
\begin{table*} 
\centering
\caption{\label{table:normal_EAKF_rmse_half} Time mean RMSEs (spreads) of the normal EAKF and the EAKF-RN in the 1/2 observation scenario, as functions of the covariance inflation factor and the half-width of covariance localization.}
\begin{tabular}{lccccc}
\hline \hline
EAKF & $l_c = 0.1$ & $l_c = 0.2$ & $l_c = 0.3$ & $l_c = 0.4$ & $l_c = 0.5$ \\
\hline
$\lambda = 1.00$ & 1.0721 (0.7049) & Div & Div & Div & Div \\
$\lambda = 1.05$ & 1.0091 (0.7457) & Div & Div & Div & Div \\
$\lambda = 1.10$ & 0.9789 (0.7868) & Div & Div & Div & Div \\
$\lambda = 1.15$ & 0.9662 (0.8209) & Div & Div & Div & Div \\
$\lambda = 1.20$ & 0.9515 (0.8566) & Div & Div & Div & Div \\
$\lambda = 1.25$ & 0.9623 (0.8929) & Div & Div & Div & Div \\
\hline \hline
EAKF-RN & $l_c = 0.1$ & $l_c = 0.2$ & $l_c = 0.3$ & $l_c = 0.4$ & $l_c = 0.5$ \\
\hline
$\lambda = 1.00$ & 1.0325 (0.7002) & 1.8256 (0.5697) & 2.1099 (0.5127) & 2.2734 (0.4736) & 2.2964 (0.4579) \\
$\lambda = 1.05$ & 1.0051 (0.7419) & 1.4072 (0.6185) & 1.9879 (0.5644) & 2.1821 (0.5269) & 2.2468 (0.5050) \\
$\lambda = 1.10$ & 0.9598 (0.7842) & 1.2313 (0.6553) & 1.8517 (0.6030) & 2.0342 (0.5699) & 2.1742 (0.5470) \\
$\lambda = 1.15$ & 0.9673 (0.8201) & 1.2024 (0.6870) & 1.6507 (0.6388) & 1.9317 (0.6015) & 2.0953 (0.5845) \\
$\lambda = 1.20$ & 0.9474 (0.8565) & 1.1788 (0.7183) & 1.5776 (0.6680) & 1.9059 (0.6336) & 2.0806 (0.6098) \\
$\lambda = 1.25$ & 0.9650 (0.8935) & 1.1856 (0.7484) & 1.5315 (0.6945) & 1.7778 (0.6603) & 2.0071 (0.6383) \\
\hline \hline
\end{tabular}
\end{table*}

\clearpage
\begin{table*} 
\centering
\caption{\label{table:normal_EAKF_rmse_quarter} As in Table \ref{table:normal_EAKF_rmse_half}, except that it is in the 1/4 observation scenario.}
\begin{tabular}{lccccc}
\hline \hline
EAKF & $l_c = 0.1$ & $l_c = 0.2$ & $l_c = 0.3$ & $l_c = 0.4$ & $l_c = 0.5$ \\
\hline
$\lambda = 1.00$ & 2.0685 (1.5730) & Div & Div & Div & Div \\
$\lambda = 1.05$ & 1.9908 (1.7849) & Div & Div & Div & Div \\
$\lambda = 1.10$ & 2.0223 (2.0447) & 2.3014 (1.5640) & Div & Div & Div \\
$\lambda = 1.15$ & 2.0819 (2.3592) & 2.2174 (1.7254) & 2.9502 (1.5820) & Div & Div \\
$\lambda = 1.20$ & 2.1903 (2.6869) & 2.1839 (1.9468) & 2.7534 (1.7191) & Div & Div \\
$\lambda = 1.25$ & 2.3586 (3.0392) & 2.2596 (2.2340) & 2.6413 (1.8780) & Div & Div \\
\hline \hline
EAKF-RN & $l_c = 0.1$ & $l_c = 0.2$ & $l_c = 0.3$ & $l_c = 0.4$ & $l_c = 0.5$ \\
\hline
$\lambda = 1.00$ & 2.0840 (1.5689) & 2.6099 (1.1984) & 3.0267 (1.0110) & 3.0453 (0.8703) & 3.0469 (0.7899) \\
$\lambda = 1.05$ & 2.0042 (1.7790) & 2.3341 (1.3762) & 2.8493 (1.1936) & 3.0573 (1.0403) & 3.1015 (0.9618) \\
$\lambda = 1.10$ & 1.9860 (2.0339) & 2.2976 (1.5332) & 2.8154 (1.3484) & 3.0527 (1.2112) & 3.1251 (1.1028) \\
$\lambda = 1.15$ & 2.0766 (2.3648) & 2.2389 (1.7244) & 2.7737 (1.4940) & 3.1247 (1.3341) & 3.2583 (1.2558) \\
$\lambda = 1.20$ & 2.1886 (2.6948) & 2.2312 (1.9710) & 2.6566 (1.6824) & 3.0992 (1.5048) & 3.2340 (1.3674) \\
$\lambda = 1.25$ & 2.3436 (3.0359) & 2.2352 (2.2344) & 2.6168 (1.8427) & 3.0977 (1.6509) & 3.2897 (1.5098) \\
\hline \hline
\end{tabular}
\end{table*}




\clearpage
\begin{figure*} 
\vspace*{2mm}
\centering
\includegraphics[width=\textwidth]{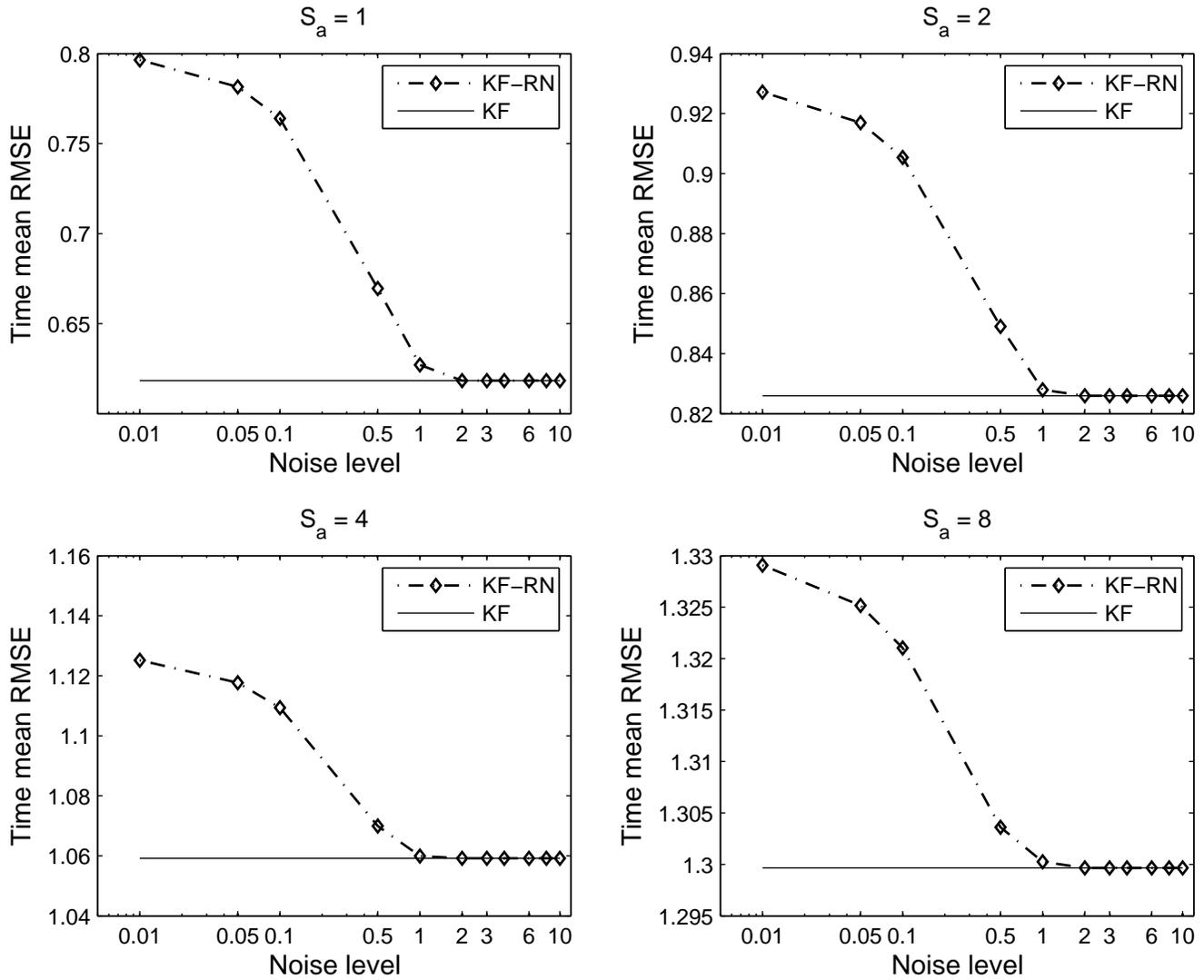}
\caption{\label{fig:KF_KFRN_rmse} Time mean RMSEs of the KF and the KF-RN as functions of the noise level coefficient in the AR1 model, with different $S_a$.}
\end{figure*}

\clearpage
\begin{figure*} 
\vspace*{2mm}
\centering
\includegraphics[width=\textwidth]{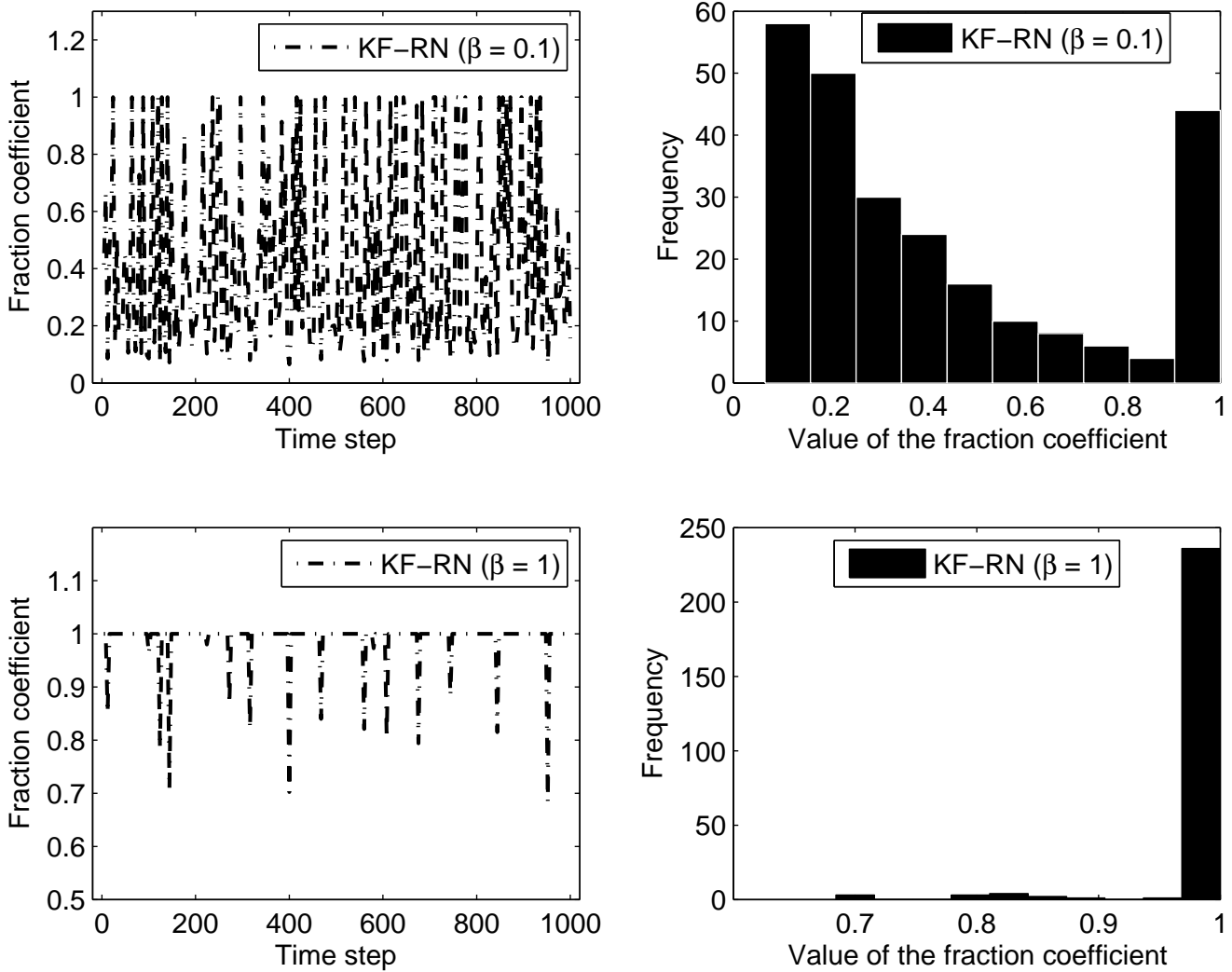}
\caption{\label{fig:KF_ts_c} Left panels: Sample time series of the fraction coefficients of the KF-RN with $\beta = 0.1$ (upper) and $\beta = 1$ (lower), respectively. Right panels: The corresponding histograms of the fraction coefficient time series.}
\end{figure*}

\clearpage
\begin{figure*} 
\vspace*{2mm}
\centering
\includegraphics[width=\textwidth]{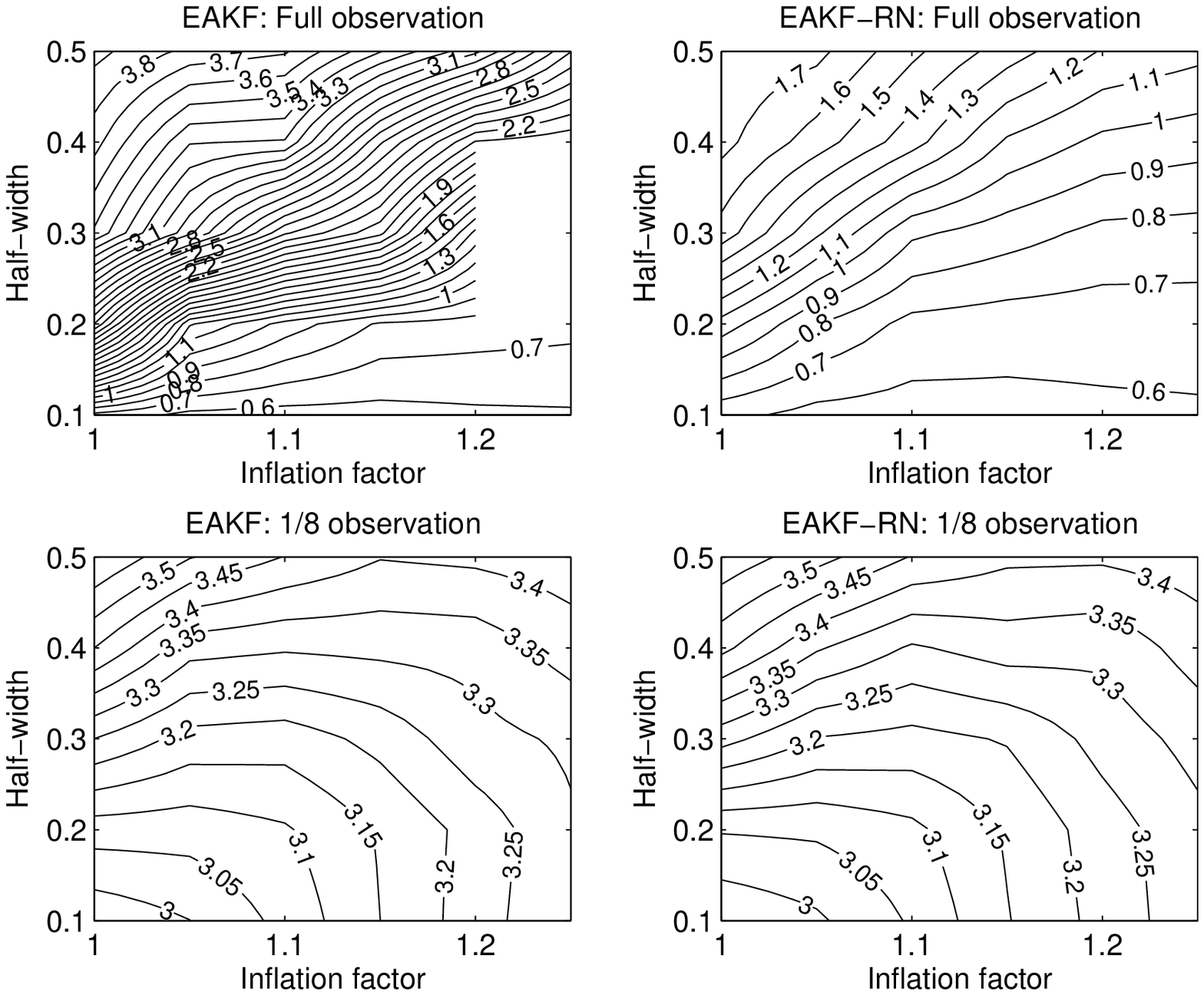}
\caption{\label{fig:normal_and_darn_EAKF_rmse} Time mean RMSEs of the normal EAKF and the EAKF-RN, as functions of inflation factor and half-width, in the full and 1/8 observation scenarios.}
\end{figure*}

\clearpage
\begin{figure*} 
\vspace*{2mm}
\centering
\includegraphics[width=\textwidth]{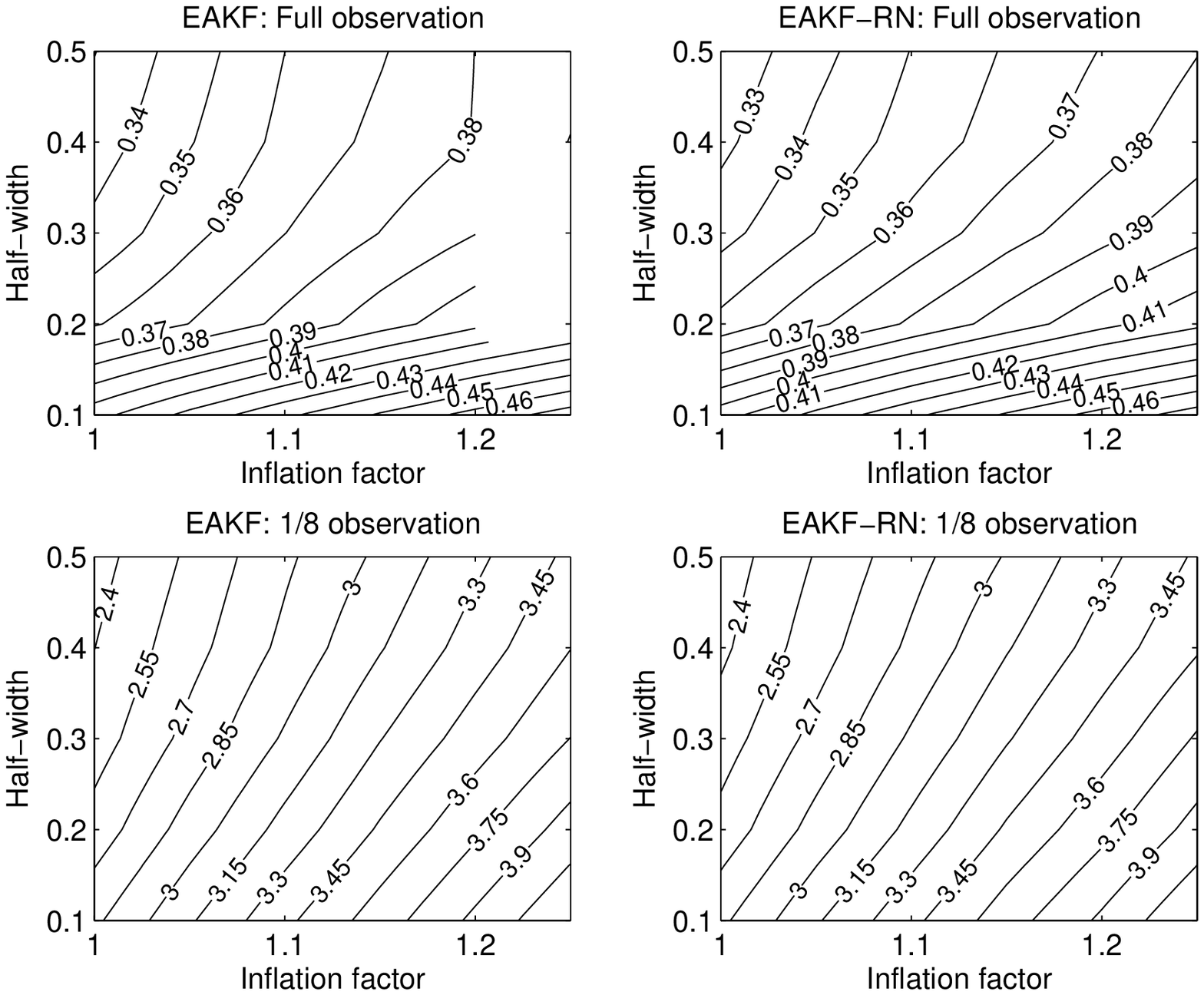}
\caption{\label{fig:normal_and_darn_EAKF_spread} Time mean spreads of the normal EAKF and the EAKF-RN, as functions of inflation factor and half-width, in the full and 1/8 observation scenarios.}
\end{figure*}

\clearpage
\clearpage
\begin{figure*} 
\vspace*{2mm}
\centering
\includegraphics[width=\textwidth]{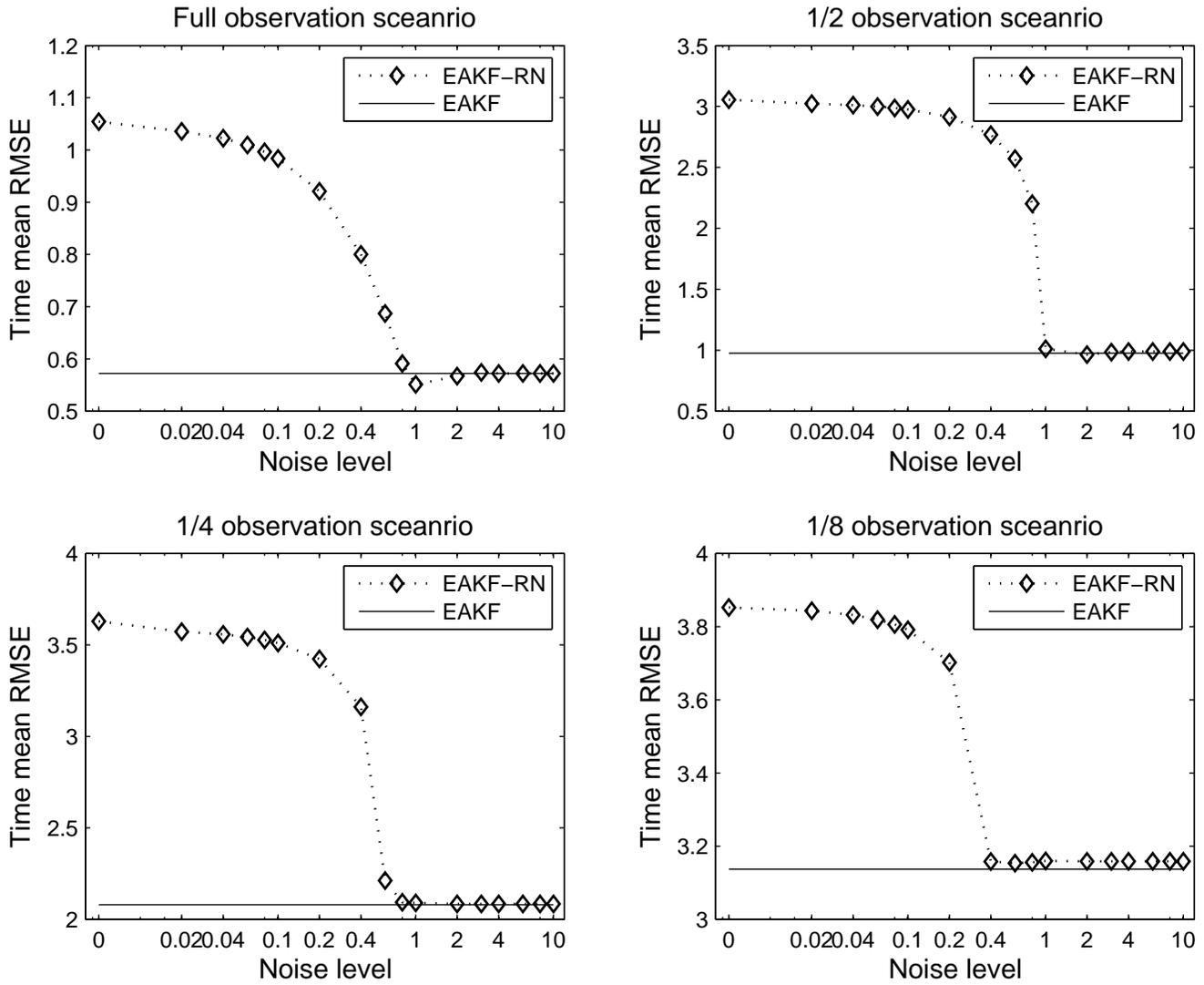}

\caption{\label{fig:EAKF_rmse_vs_noise_level_obsOP_lc01_delta015} Time mean RMSEs of the normal EAKF and the EAKF-RN as functions of the noise level coefficient in different observation scenarios, with $\lambda = 1.15$ and $l_c = 0.1$.}
\end{figure*}

\clearpage
\clearpage
\begin{figure*} 
\vspace*{2mm}
\centering
\includegraphics[width=\textwidth]{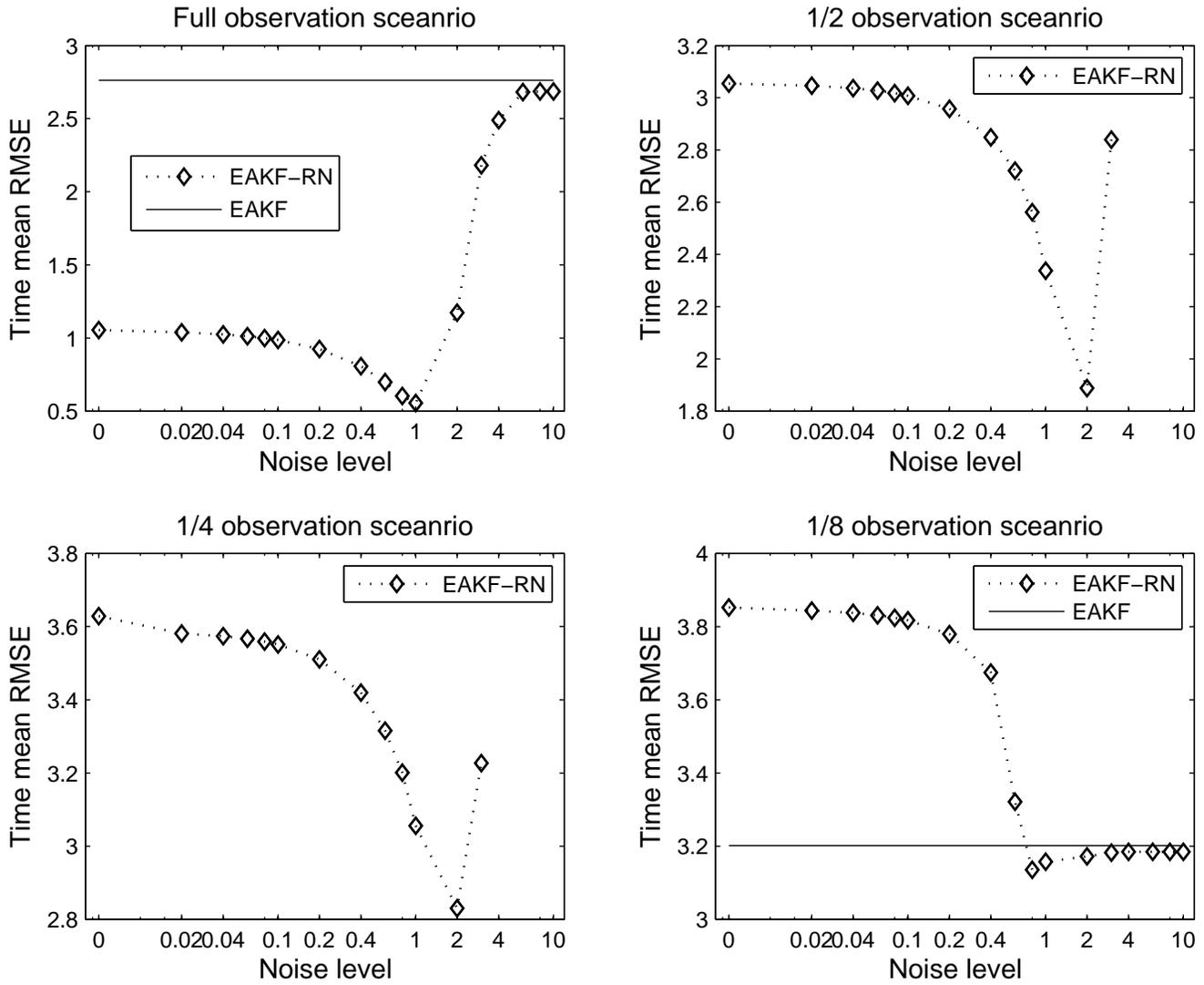}

\caption{\label{fig:EAKF_rmse_vs_noise_level_obsOP_lc03_delta005} As in Fig. \ref{fig:EAKF_rmse_vs_noise_level_obsOP_lc01_delta015}, but with $\lambda = 1.05$ and $l_c = 0.3$ for both the filters. Note that in the 1/2 and 1/4 observation scenarios divergences of the normal EAKF are spotted, hence no horizontal lines are indicated in the corresponding plots. The EAKF-RN also diverges in the 1/2 and 1/4 observation scenarios for $\beta \geq 4$.}
\end{figure*}

\clearpage
\clearpage
\begin{figure*} 
\vspace*{2mm}
\centering
\includegraphics[width=\textwidth]{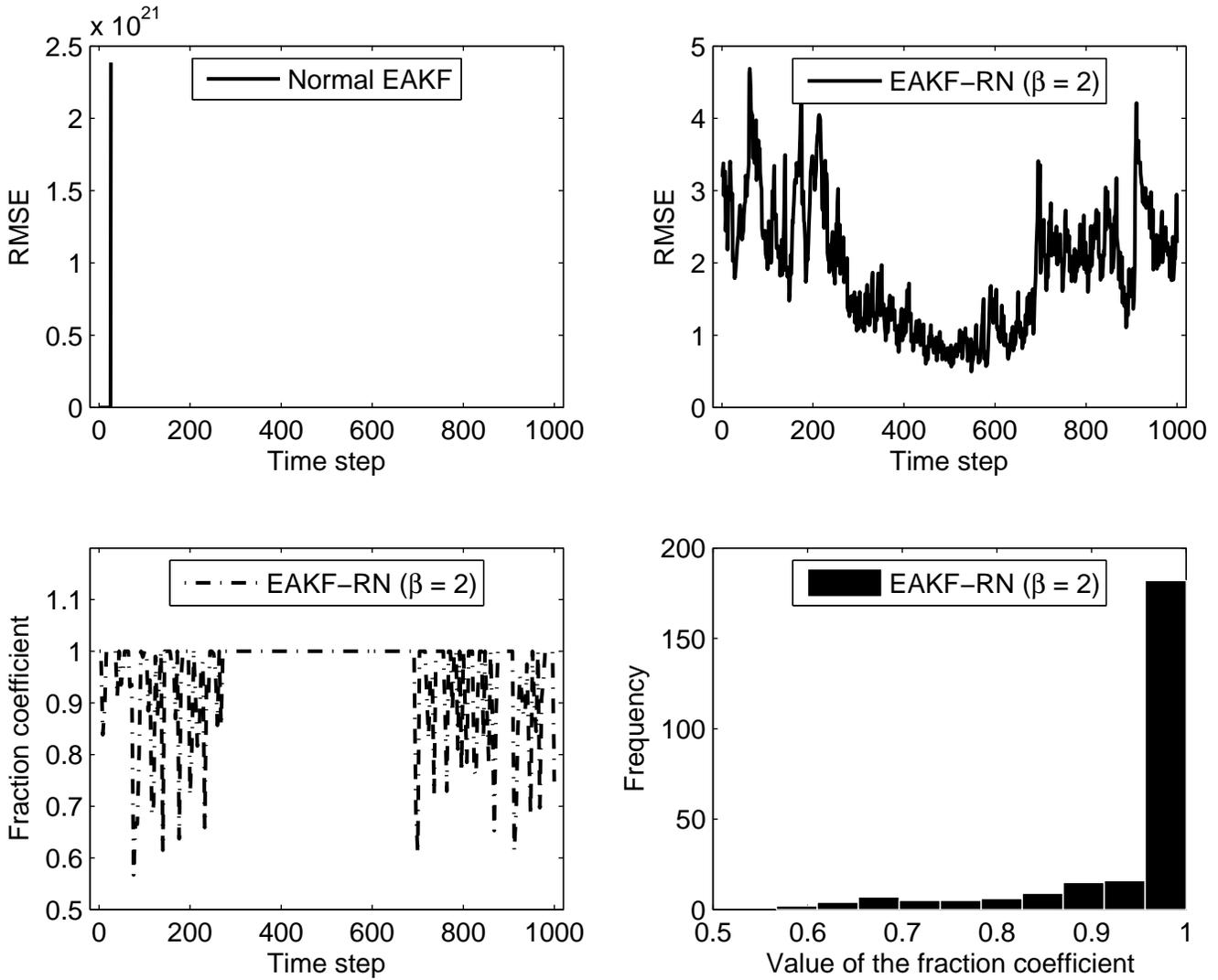}

\caption{\label{fig:EAKF_ts_c} Upper left: sample time series of the RMSE of the normal EAKF in the 1/2 observation scenario; Upper right: sample time series of the RMSE of the EAKF-RN ($\beta = 2$) under the same experiment settings as the EAKF; Lower left: corresponding fraction coefficient $c_k$ in the EAKF-RN ($\beta = 2$); Lower right: corresponding histogram of $c_k$.}
\end{figure*}

\clearpage
\clearpage
\begin{figure*} 
\vspace*{2mm}
\centering
\includegraphics[width=\textwidth]{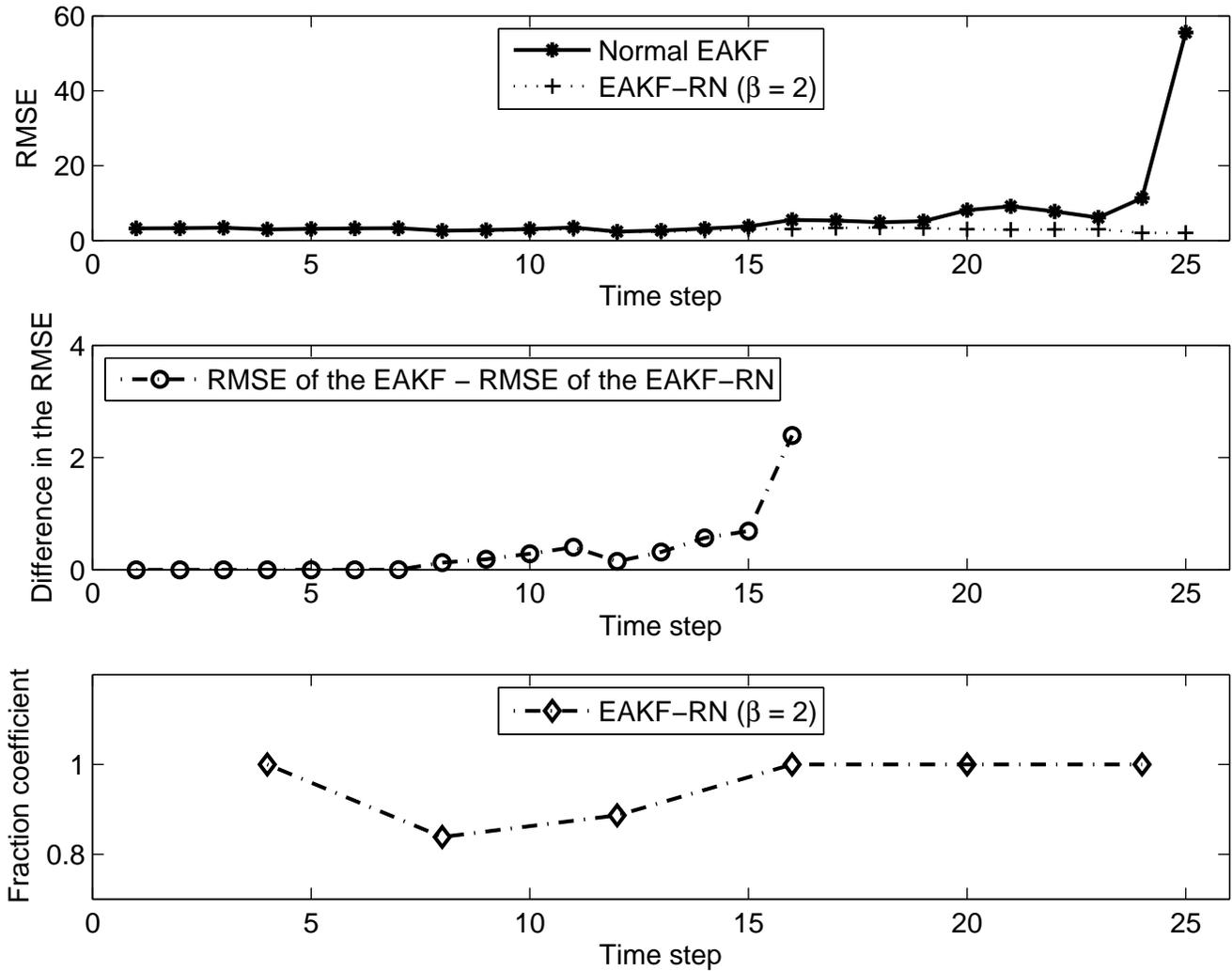}

\caption{\label{fig:EAKF_ts_c_interval} Upper: the RMSE of the EAKF (solid line with asterisks) and EAKF-RN ($\beta = 2$, dotted line with plus signs) between the time instant $k=1$ and $k=25$; Middle: difference in the RMSE ( = RMSE of the EAKF - RMSE of the EAKF-RN) between $k = 1$ and $k = 16$; Lower: the fraction coefficient of the EAKF-RN ($\beta = 2$) between $k = 1$ and $k = 25$.}
\end{figure*}

\clearpage
\begin{figure*} 
\vspace*{2mm}
\centering
\includegraphics[width=\textwidth]
{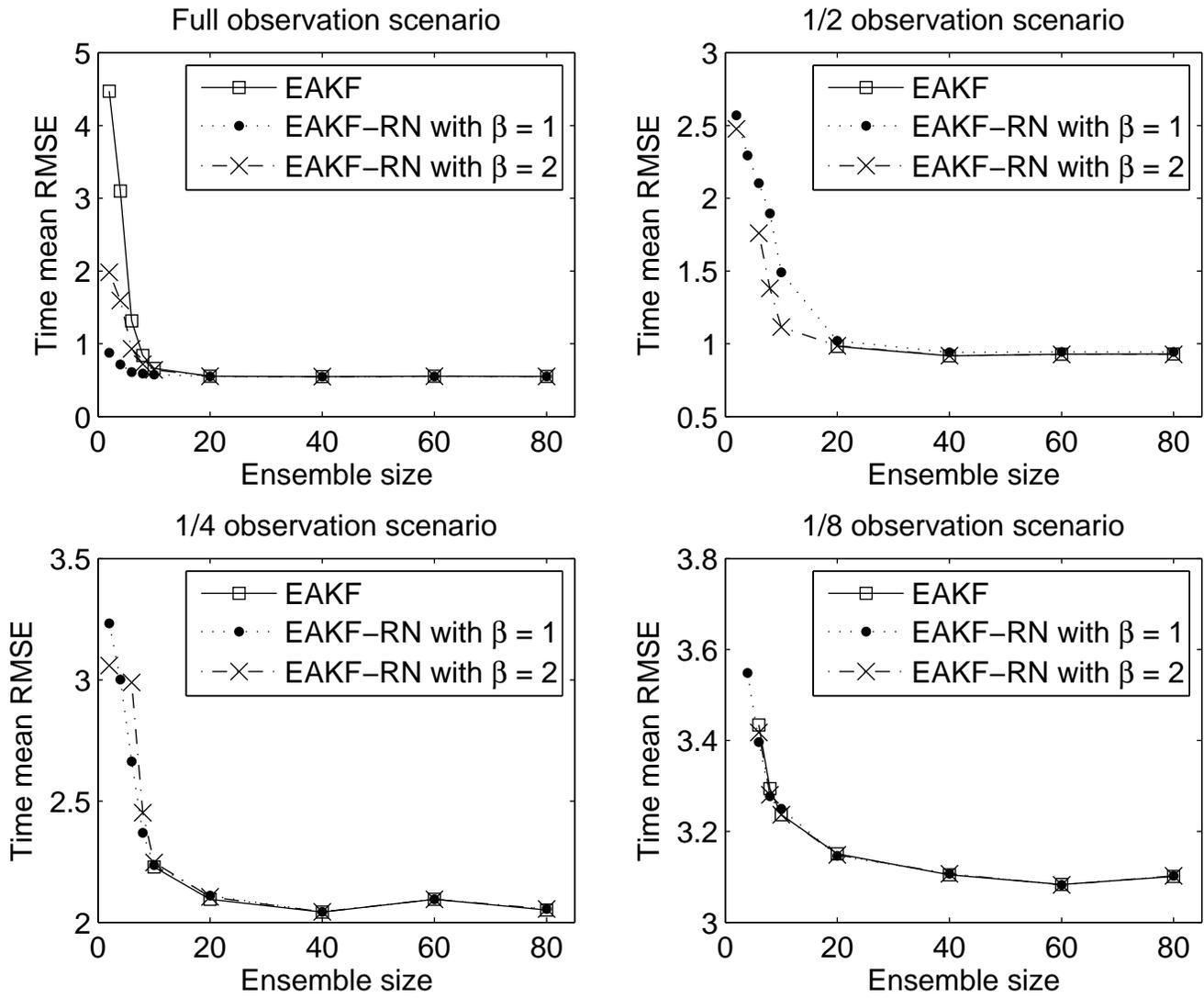}
\caption{\label{fig:EAKF_RN_rmse_vs_ensize} Time mean RMSEs of the EAKF and the EAKF-RN, as functions of the ensemble size in different observation scenarios.}
\end{figure*}

\clearpage
\begin{figure*} 
\vspace*{2mm}
\centering
	
\includegraphics[width=\textwidth]{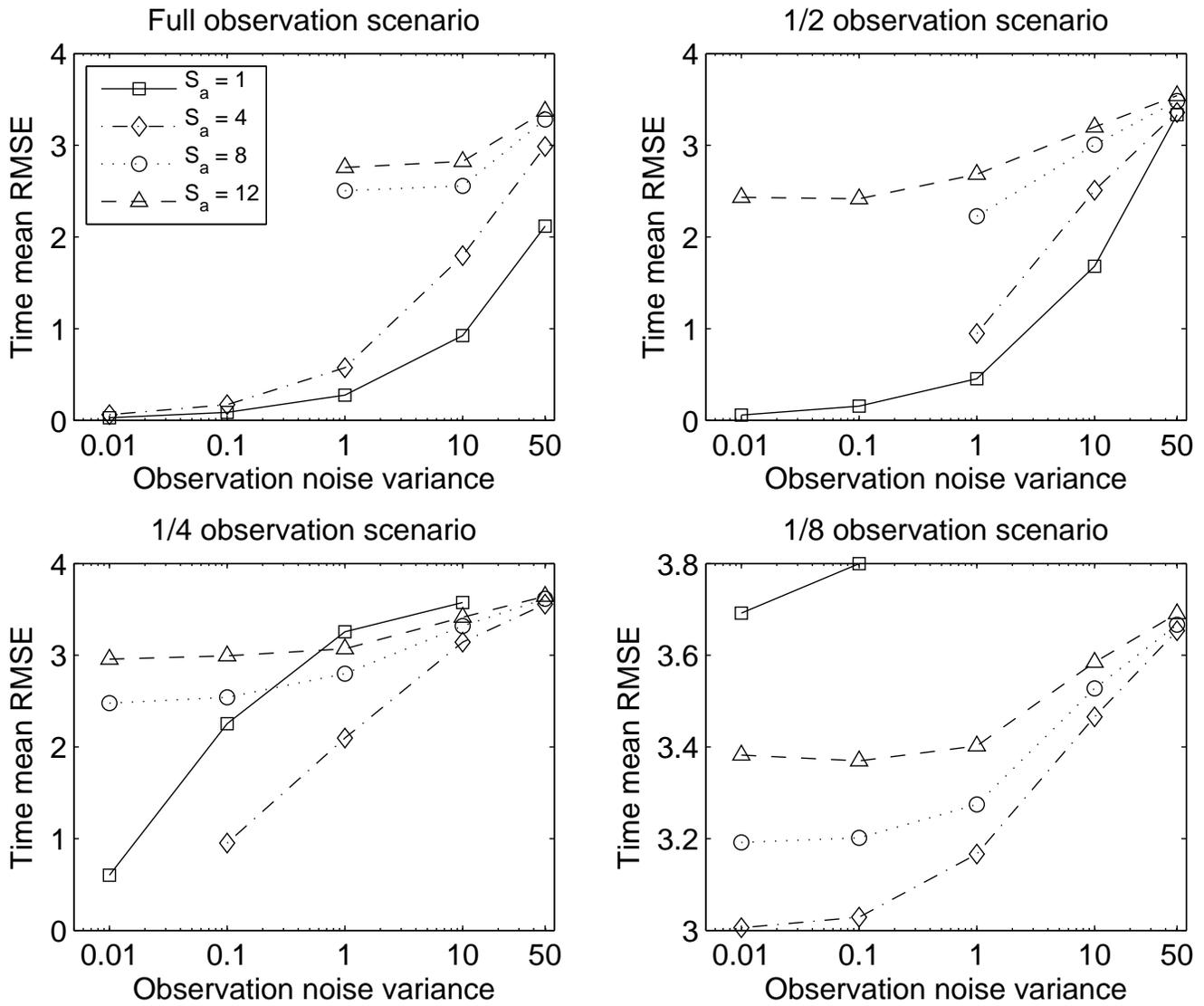}

\caption{\label{fig:normal_rmse_varying_amStep_obsLvl} Time mean RMSEs of the normal EAKF, as functions of the assimilation step $S_a$ and the observation noise variance, in different observation scenarios.}
\end{figure*}

\clearpage
\begin{figure*} 
\vspace*{2mm}
\centering
	
\includegraphics[width=\textwidth]{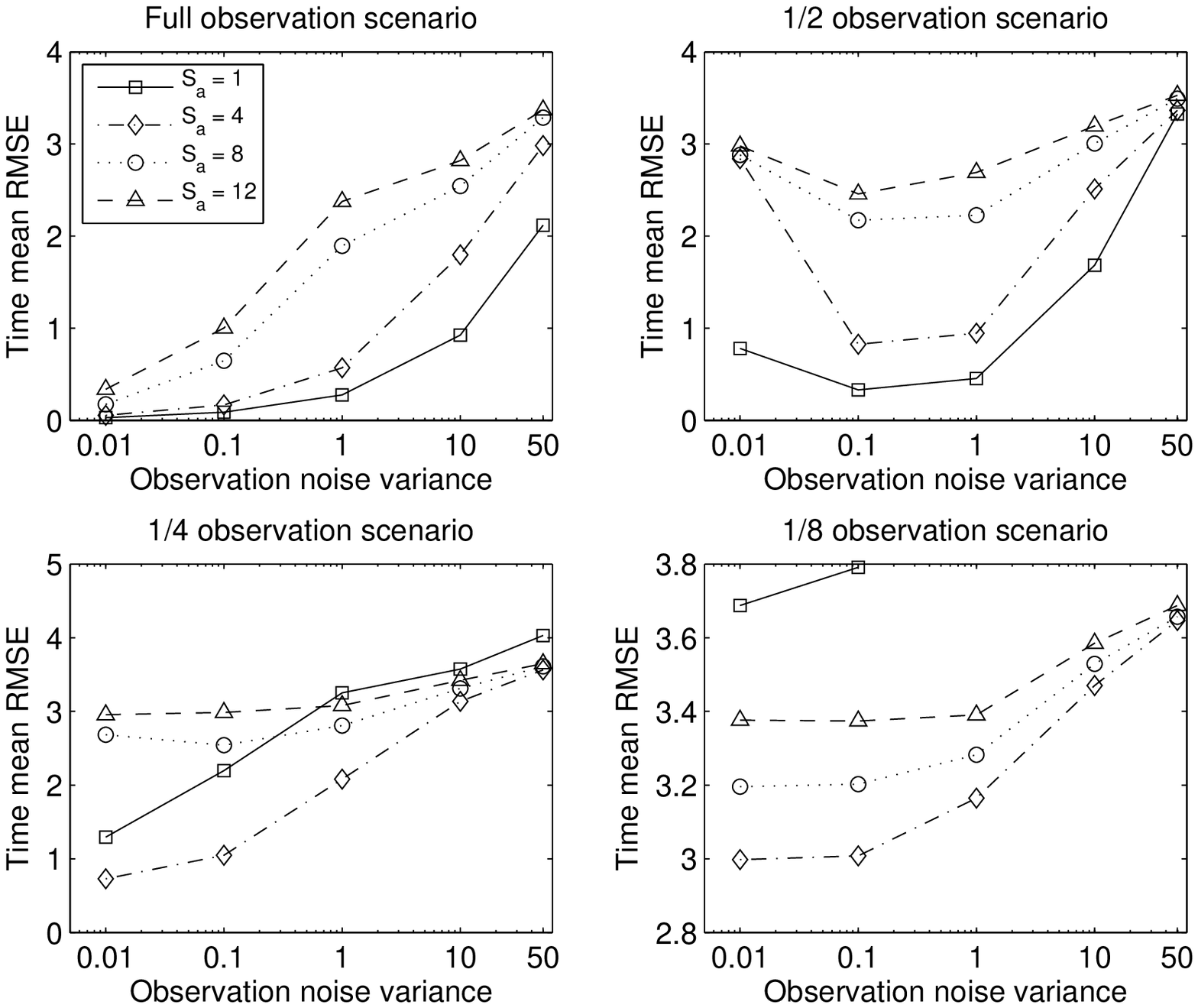}

\caption{\label{fig:darn_rmse_varying_amStep_obsLvl} As in Fig. \ref{fig:normal_rmse_varying_amStep_obsLvl}, but for the EAKF-RN with $\beta = 2$.}
\end{figure*}

\clearpage
\begin{figure*} 
\vspace*{2mm}
\centering
	
\includegraphics[width=\textwidth]{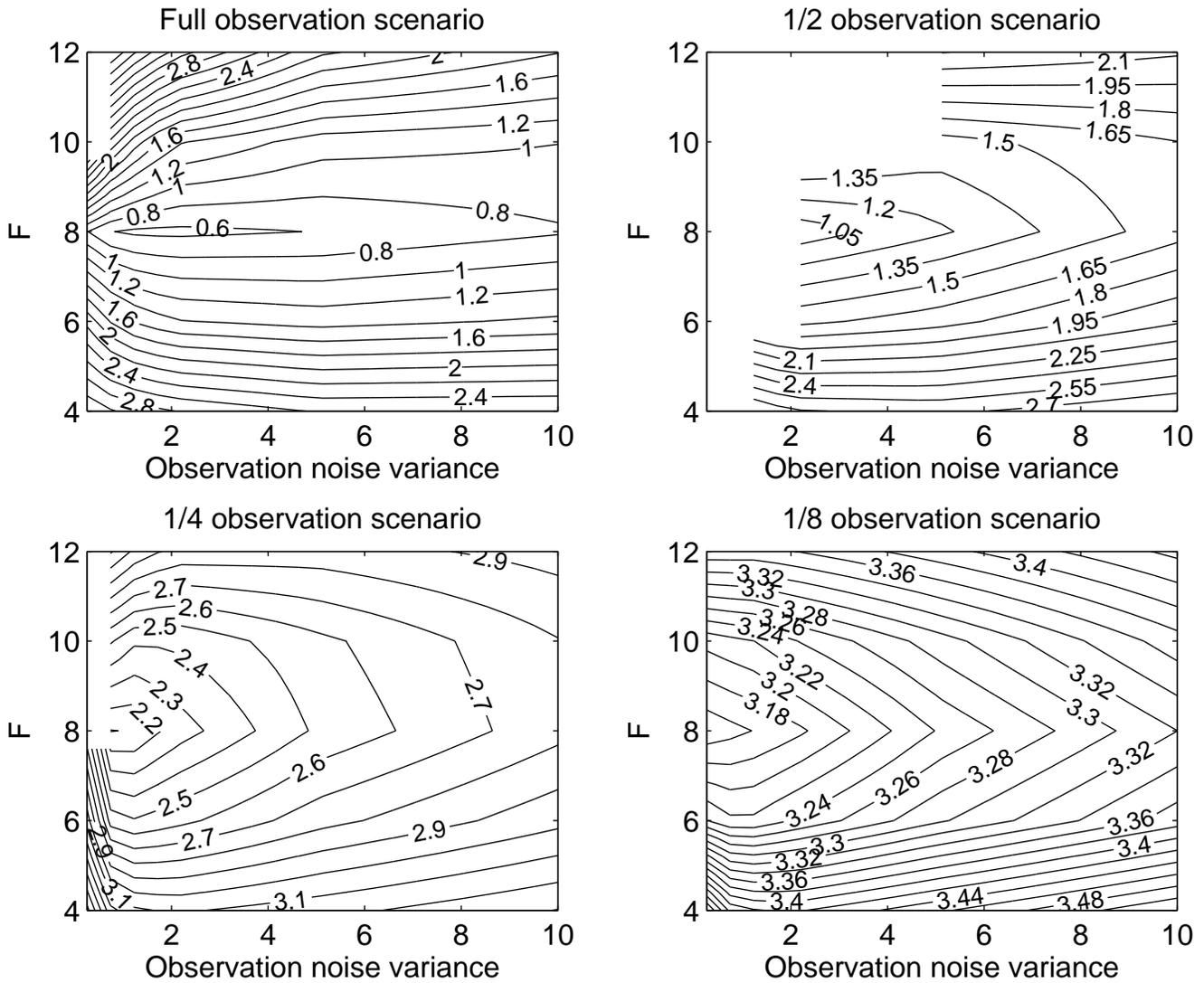}

\caption{\label{fig:normal_rmse_varying_F_and_c} Time mean RMSEs of the EAKF, as functions of the (possibly) mis-specified driving force F and the observation noise variance $\gamma$, in different observation scenarios.}
\end{figure*}

\clearpage
\begin{figure*} 
\vspace*{2mm}
\centering
	
\includegraphics[width=\textwidth]{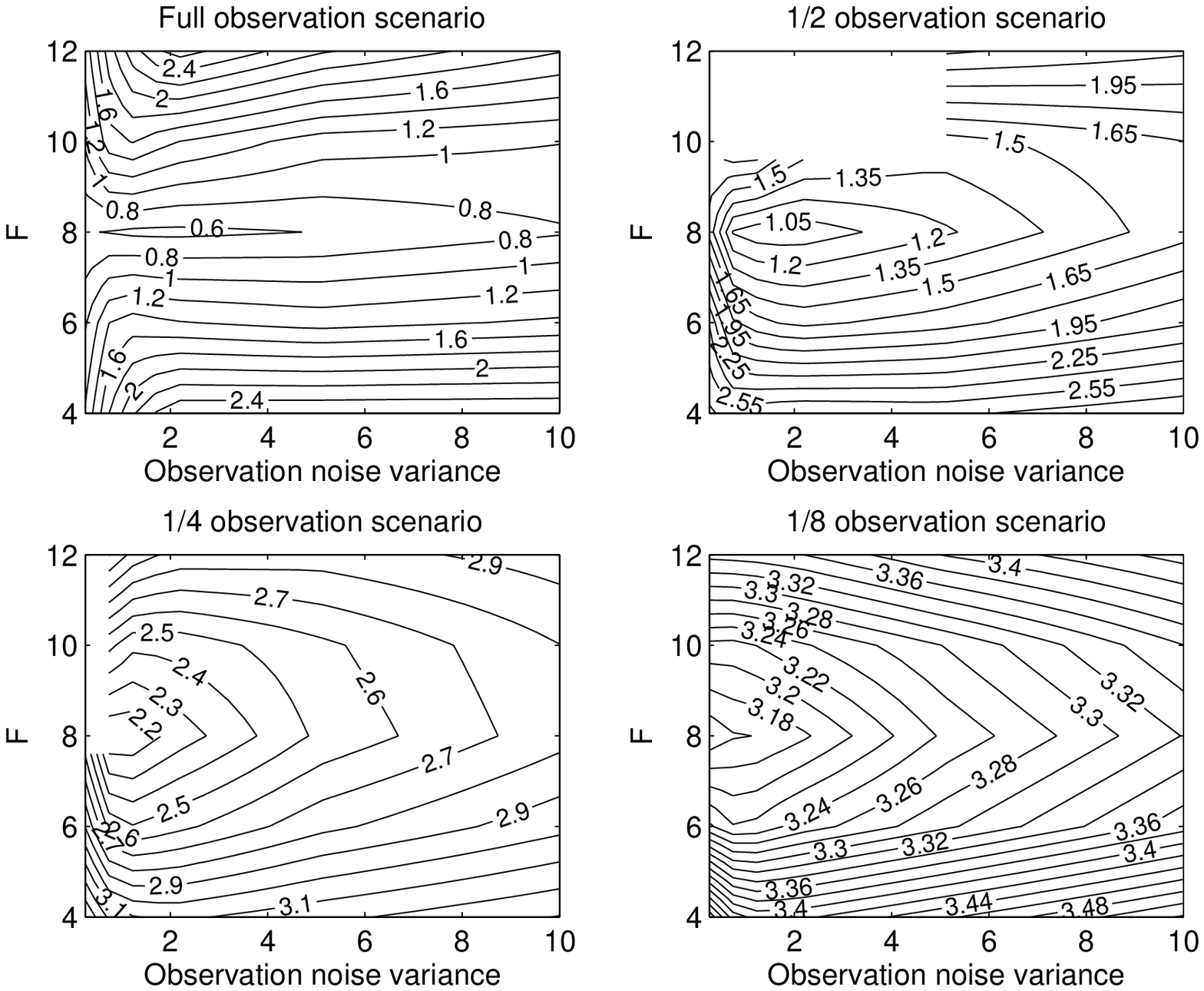}

\caption{\label{fig:darn_rmse_varying_F_and_c} As in Fig. \ref{fig:normal_rmse_varying_F_and_c}, but for the EAKF-RN with $\beta = 2$.}
\end{figure*}

\end{document}